\documentclass[prl,12pt]{revtex4}
\usepackage{amssymb}
\usepackage{amsmath}
\usepackage{graphicx}
\usepackage{color}
\DeclareGraphicsExtensions{.jpg,.jpeg,.pdf,.png,.mps,.eps,.ps}

\newcommand{\beq}{\begin{equation}}
\newcommand{\eeq}{\end{equation}}
\newcommand{\bea}{\begin{eqnarray}}
\newcommand{\ena}{\end{eqnarray}}
\newcommand{\etal}{{\it et al.}}
\newcommand{\ie}{{\it i.e.}}
\newcommand{\eg}{{\it e.g.}}
\newcommand{\etc}{{\it etc.}}
\newcommand{\lsim}{\mathrel{\mathop{\kern 0pt \rlap
{\raise.2ex\hbox{$<$}}}
\lower.9ex\hbox{\kern-.190em $\sim$}}}
\newcommand{\gsim}{\mathrel{\mathop{\kern 0pt \rlap
{\raise.2ex\hbox{$>$}}}
\lower.9ex\hbox{\kern-.190em $\sim$}}}

\newcommand{\physics}[1]{{\tt physics/#1}}

\newcommand{\plb}[3]{Phys.\ Lett.\ B\ {\bf #1}, #3 (#2)}

\newcommand{\pr}[3]{Phys.\ Rev.\ {\bf #1}, #3 (#2)}

\renewcommand{\rmp}[3]{Rev.\ Mod.\ Phys.\ {\bf #1}, #3 (#2)}

\newcommand{\href}[2]{#1}

\definecolor{cyan}{cmyk}{1.,0.,0.,0.5}
\definecolor{magenta}{cmyk}{0.,1.,0.,0.5}
\definecolor{verdatre}{cmyk}{0.5,0.,0.5,0.5}
\definecolor{yellow}{cmyk}{0.,0.,0.2,0.0}
\definecolor{rouge}{cmyk}{0.,0.4,0.6,0.0}
\definecolor{orange}{cmyk}{0.,0.5,0.5,0.}
\definecolor{violet}{rgb}{0.5,0.,0.5}

\begin{document}

\noindent
\title{Viewing the CPT Invariance as a Basic Postulate in Physics}
 \vskip 1.cm
\author{Guang-jiong Ni $^{\rm a,b}$}
\email{\ pdx01018@pdx.edu}
\affiliation{$^{\rm a}$ Department of Physics, Portland State University, Portland, OR97207, U. S. A.\\
$^{\rm b}$ Department of Physics, Fudan University, Shanghai, 200433, China}

\author{Suqing Chen $^{\rm b}$}
\email{\ suqing\_chen@yahoo.com}
\affiliation{$^{\rm b}$ Department of Physics, Fudan University, Shanghai, 200433, China}

\author{Jianjun Xu $^{\rm b}$}
\email{\ xujj@fudan.edu.cn}
\affiliation{$^{\rm b}$ Department of Physics, Fudan University, Shanghai, 200433, China}

\vskip 0.5cm
\date{\today}

\vskip 0.5cm
\begin{abstract}
The $CPT$ invariance has been firmly established via experimental tests. Its theoretical implication and derivation at two levels of both relativistic quantum mechanics ($RQM$) and quantum field theory ($QFT$) are discussed. Being a basic symmetry, the $CPT$ invariance can be expressed as ${\cal P}{\cal T}={\cal C}$, where ${\cal P}{\cal T}$ represent the "strong reflection", \ie, the (newly defined) space-time inversion (${\bf x}\to -{\bf x}, t\to-t$), invented by L\"{u}ders and Pauli in the proof of the $CPT$ theorem and ${\cal C}$ the new particle-antiparticle transformation proposed by Lee and Wu. Actually, the renamed $CPT$ invariance, ${\cal P}{\cal T}={\cal C}$, could be viewed as a basic postulate being injected implicitly into the theory since the nonrelativistic quantum mechanics ($NRQM$) was combined with the theory of special relativity ($SR$) to become $RQM$ and then the $QFT$. The Klein-Gordon ($KG$) equation is highlighted to become a self-consistent theory in $RQM$, based on two sets of wavefunctions ($WFs$) and momentum-energy operators for particle and  antiparticle respectively, together with the above postulate. Hence the Klein paradox for both $KG$ equation and Dirac equation can be solved without resorting to the "hole theory".  \\
{\bf Keywords}:\;CPT invariance, Antiparticle, Quantum mechanics, Quantum field theory\\
{\bf PACS}:\; 03.65.-w; 03.65.Ta; 03.65.Ud; 11.10.-z

\end{abstract}

\maketitle \vskip 1cm

\section{I. Introduction}
\label{sec:introduction}

\vskip 0.1cm

The famous paper by Einstein, Podolsky and Rosen (EPR \cite{1}) in 1935 is not easy to read, at least to us. And the deep implication of a remarkable $EPR$ experiment by CPLEAR collaboration on $K^0-\bar{K}^0$ system in 1998 \cite{2} is not easy to explore either. We discuss these two papers in parallel in section II before we will be able to extract some new conception about the antiparticle's wavefunction ($WF$) and momentum-energy operators. Then in section III the Klein-Gordon ($KG$) equation is highlighted to become a self-consistent theory in $RQM$ based on two sets of wavefunctions ($WFs$) and momentum-energy operators for particle and antiparticle respectively together with a symmetry under the (newly defined) space-time inversion ${\cal P}{\cal T}$ (${\bf x}\to -{\bf x}, t\to-t$) or mass inversion ($m\to-m$) which is inspired by the Feshbach-Villars dissociation of KG equation (1958, \cite{25}). The above symmetry is further discussed and is identified with the "strong reflection" invariance invented by L\"{u}ders and Pauli in their proof of the $CPT$ theorem (1954-1957, \cite{31,32,33}) in QFT as well as the new definition (${\cal C}$) for particle-antiparticle transformation proposed by Lee and Wu (1965, \cite{21}). All discussions are combined to become a renamed symmetry for the $CPT$ invariance as ${\cal P}{\cal T}={\cal C}$. In section IV, the Dirac equation is discussed. In section V, we discuss $QFT$. Section VI  is a brief summary and discussion. The Klein paradox is solved in the Appendix for both $KG$ equation and Dirac equation without resorting to the "hole theory".

\section{II. What the $K^0\bar{K}^0$ correlation experimental data are telling?}
\label{sec:correlation}
\setcounter{equation}{0}
\renewcommand{\theequation}{2.\arabic{equation}}
\vskip 0.1cm

To our knowledge, beginning from Bohm \cite{3} and Bell \cite{4}, physicists gradually turned their research of EPR paradox onto the entangled state composed of electrons, especially photons with spin and achieved fruitful results. However, as pointed out by Guan (1935-2007), EPR's paper \cite{1} is focused on two spinless particles and Guan found that there is a commutation relation hiding in such a system as follows \cite{5}:\\
Consider two particles in one dimensional space with positions $x_i\,(i=1,2)$ and momentum operators $\hat{p}_i=-i\hbar\frac{\partial}{\partial x_i}
$. Then a commutation relation arises as
\begin{equation}\label{1}
[x_1-x_2,\hat{p}_1+\hat{p}_2]=0
\end{equation}
According to QM's principle, there may be a kind of common eigenstate having eigenvalues of these two commutative (\ie, compatible)observables like:
\begin{equation}\label{2}
p_1+p_2=0,\;(p_2=-p_1)\quad \text{and}\quad (x_1-x_2)=D
\end{equation}
with $D$ being their distance. The existence of such kind of eigenstate described by Eq.(\ref{2}) puzzled Guan, he asked: "How can such kind of quantum state be realized?" A discussion between Guan and one of present authors (Ni) in 1998 led to a paper \cite{6}. \\
Here we are going to discuss further, showing that the correlation experiment on a $K^0\bar{K}^0$ system (which just realized an entangled state composed of two spinless particles) in 1998 by CPLEAR collaboration \cite{2} actually revealed some important features of QM and then answered the puzzle raised by EPR in a surprising way. First, besides Eq.(\ref{1}), let us consider another three commutation relations simultaneously:
\begin{equation}\label{3}
[t_1+t_2,\hat{E}_1-\hat{E}_2]=0
\end{equation}

\begin{equation}\label{4}
 [x_1+x_2,\hat{p}_1-\hat{p}_2] =0
\end{equation}

\begin{equation}\label{5}
 [t_1-t_2,\hat{E}_1+\hat{E}_2]=0
\end{equation}
($E_i=i\hbar\frac{\partial}{\partial t_i}$ with $t_i$ being the time during which the {\it i}'th particle is detected). In accordance with Ref.\cite{2}, we also focus on back-to-back events. The evolution of $K^0\bar{K}^0$'s wavefunction (WF) will be considered in three inertial frames: The center-of-mass system $S$ is at rest in laboratory with its origin $x=0$ located at the apparatus' center, where the antiprotons' beam is stopped inside a hydrogen gas target to create $K^0\bar{K}^0$ pairs by $p\bar{p}$ annihilation. The $K^0\bar{K}^0$ pairs are detected by a cylindrical tracking detector located inside a solenoid providing a magnetic field parallel to the antiprotons' beam. For back-to-back events, the space-time coordinates in Eqs.(\ref{1})-(\ref{5}) refer to particles moving to the right ($x_1>0$) and left ($x_2<0$) respectively. Second, we take an inertial system $S'$ with its origin located at particle 1 (\ie, $x'_1=0$). $S'$ is moving in a uniform velocity $v$ with respect to $S$. (For Kaon's momentum of $800\,MeV/c,\;\beta=v/c=0.849$). Another $S''$ system is chosen with its origin located at particle 2 ($x''_2=0$). $S''$ is moving in a velocity ($-v$) with respect to $S$. Thus we have Lorentz transformation among the space-time coordinates being
\begin{equation}\label{6}
\left\{
  \begin{array}{ll}
    x'=\dfrac{x-vt}{\sqrt{1-\beta^2}}, & \\[5mm]
    t'=\dfrac{t-vx/c^2}{\sqrt{1-\beta^2}}, &
  \end{array}
\right.\qquad
\left\{
  \begin{array}{ll}
    x''=\dfrac{x+vt}{\sqrt{1-\beta^2}}, & \\[5mm]
    t''=\dfrac{t+vx/c^2}{\sqrt{1-\beta^2}}, &
  \end{array}\right.
\end{equation}
Here $t'_1$ and $t''_2$ correspond to the proper time $t_a$ and $t_b$ in Ref.\cite{2} respectively. The common time origin $t=t'=t''=0$ is adopted.\\
A $K^0\bar{K}^0$ pair, created in a $J^{PC}=1^{--}$ antisymmetric state, can be described by a two-body WF depending on time as (\cite{2}, see also \cite{7,8})
\begin{equation}\label{7}\begin{array}{l}
|\Psi(0,0)\rangle^{(antisym)}=\dfrac{1}{\sqrt2}\left[|K^0(0)\rangle_a|\bar{K}^0(0)\rangle_b
-|\bar{K}^0(0)\rangle_a|K^0(0)\rangle_b\right]\\[5mm]
|\Psi(t_a,t_b)\rangle^{(antisym)}=\dfrac{1}{\sqrt2}\left[|K_S(0)\rangle_a|K_L(0)\rangle_be^{-i(\alpha_St_a+\alpha_Lt_b)}
-|K_L(0)\rangle_a|K_S(0)\rangle_be^{-i(\alpha_Lt_a+\alpha_St_b)}\right]
\end{array}\end{equation}
with
\begin{equation}\label{7b}
|K_S\rangle=\dfrac{1}{\sqrt2}[|K^0\rangle-|\bar{K}^0\rangle],\;|K_L\rangle=\dfrac{1}{\sqrt2}[|K^0\rangle+|\bar{K}^0\rangle]
\end{equation}
where the CP violation has been neglected and $\alpha_{S,L}=m_{S,L}-i\gamma_{S,L}/2$, $m_{S,L}$ and $\gamma_{S,L}$ being the $K_{S,L}$ masses and decay widths, respectively. From Eq.(\ref{7}), the intensities of events with like-strangeness ($K^0K^0$ or $\bar{K}^0\bar{K}^0$) and unlike-strangeness ($K^0\bar{K}^0$ or $\bar{K}^0K^0$) can be evaluated as
\begin{equation}\label{8}
I_{like}^{(antisy)}(t_a,t_b)=\dfrac{1}{8}e^{-2\gamma\tilde{t}}\left\{e^{-\gamma_S|t_a-t_b|}+e^{-\gamma_L|t_a-t_b|}
-2e^{-\gamma|t_a-t_b|}\cos[\Delta m(t_a-t_b)]\right\}
\end{equation}

\begin{equation}\label{9}
I_{unlike}^{(antisy)}(t_a,t_b)=\dfrac{1}{8}e^{-2\gamma\tilde{t}}\left\{e^{-\gamma_S|t_a-t_b|}+e^{-\gamma_L|t_a-t_b|}
+2e^{-\gamma|t_a-t_b|}\cos[\Delta m(t_a-t_b)]\right\}
\end{equation}
where $\Delta m=m_L-m_S,\;\gamma=(\gamma_S+\gamma_L)/2$ and $\tilde{t}=t_a\,(\text{for}\;t_a<t_b)$ or $\tilde{t}=t_b\,(\text{for}\;t_a>t_b)$.\\
Similarly, for $K^0\bar{K}^0$ created in a $J^{PC}=0^{++}$ or $2^{++}$ symmetric state as:
\begin{equation}\label{10}\begin{array}{l}
|\Psi(0,0)\rangle^{(sym)}=\dfrac{1}{\sqrt2}\left[|K^0(0)\rangle_a|\bar{K}^0(0)\rangle_b
+|\bar{K}^0(0)\rangle_a|K^0(0)\rangle_b\right]\\[5mm]
|\Psi(t_a,t_b)\rangle^{(sym)}=\dfrac{1}{\sqrt2}\left[|K_L(0)\rangle_a|K_L(0)\rangle_be^{-i(\alpha_Lt_a+\alpha_Lt_b)}
-|K_S(0)\rangle_a|K_S(0)\rangle_be^{-i(\alpha_St_a+\alpha_St_b)}\right]
\end{array}\end{equation}
the predicted intensities read
\begin{equation}\label{11}\begin{array}{l}
I_{like}^{(sym)}(t_a,t_b)=\dfrac{1}{8}\left\{e^{-\gamma_S(t_a+t_b)}+e^{-\gamma_L(t_a+t_b)}
-2e^{-\gamma(t_a+t_b)}\cos[\Delta m(t_a+t_b)]\right\}\\[5mm]
I_{unlike}^{(sym)}(t_a,t_b)=\dfrac{1}{8}\left\{e^{-\gamma_S(t_a+t_b)}+e^{-\gamma_L(t_a+t_b)}
+2e^{-\gamma(t_a+t_b)}\cos[\Delta m(t_a+t_b)]\right\}
\end{array}\end{equation}
The experiment \cite{2} reveals that the $K^0\bar{K}^0$ pairs are mainly created in the antisymmetric state shown by Eqs.(\ref{8})-(\ref{9}) while the contribution in a symmetric state shown by Eqs.(\ref{10})-(\ref{11}) accounts for $7.4\%$.\\
What we learn from Ref.\cite{2} in combination with Eqs.(\ref{1})-(\ref{5}) are as follows:\\
(a)\ Because only back-to-back events are involved in the $S$ system, we denote three commutative operators as: the "distance" operator $\hat{D}=x_1-x_2=v(t_1+t_2)$, $\hat{A}=\hat{p}_1+\hat{p}_2$ and $\hat{B}=\hat{E}_1-\hat{E}_2$, Eqs.(\ref{1}) and (\ref{3}) read
\begin{equation}\label{12}
[\hat{D},\hat{A}]=0,\;[\hat{D},\hat{B}]=0,\;[\hat{A},\hat{B}]=0
\end{equation}
So they may have a kind of common eigenstate during the measurement composed of $K^0K^0$ and projected from the symmetric state shown by Eq.(\ref{10}). It is assigned by a continuous eigenvalue $D_j=v(t_1+t_2)$ (with continuous index $j$) of operator $\hat{D}$ acting on the WF, $\Psi^{sym}_{K^0K^0}(x_1,t_1;x_2,t_2)$, as\footnotemark[1]\footnotetext[1]{The WF reads approximately as:
\begin{equation*}
\Psi^{sym}_{K^0K^0}(x_1,t_1;x_2,t_2)\sim e^{i(p_1x_1-E_1t_1)}e^{i(p_2x_2-E_2t_2)}\eqno{(2.14b)}
\end{equation*}
which can be calculated from $\langle K^0K^0|\Psi(t_a,t_b)\rangle^{sym}$ with two terms. The squares of WF's  amplitude reproduces the $I_{like}^{(sym)}(t_a,t_b)$ in Eq.(\ref{11}).}
\begin{equation*}\label{13}
\hat{D}\Psi^{sym}_{K^0K^0}(x_1,t_1;x_2,t_2)=D_j\Psi^{sym}_{K^0K^0}(x_1,t_1;x_2,t_2)
=v(t_1+t_2)\Psi^{sym}_{K^0K^0}(x_1,t_1;x_2,t_2)\eqno{(2.14a)}
\end{equation*}
\setcounter{equation}{14}

\begin{equation}\label{14}
\hat{A}\Psi^{sym}_{K^0K^0}(x_1,t_1;x_2,t_2)=A^{like}_j\Psi^{sym}_{K^0K^0}(x_1,t_1;x_2,t_2)=(p_1+p_2)\Psi^{sym}_{K^0K^0}(x_1,t_1;x_2,t_2)
\end{equation}

\begin{equation}\label{15}
\hat{B}\Psi^{sym}_{K^0K^0}(x_1,t_1;x_2,t_2)=B^{like}_j\Psi^{sym}_{K^0K^0}(x_1,t_1;x_2,t_2)=(E_1-E_2)\Psi^{sym}_{K^0K^0}(x_1,t_1;x_2,t_2)
\end{equation}
where the lowest eigenvalue of $\hat{A}$ is $A^{like}_j=p_1+p_2=0,\,(p_2=-p_1)$, and that of $\hat{B}$ is $B^{like}_j=E_1-E_2=0,\,(E_2=E_1)$ respectively. These eigenstates of like-strangeness events predicted by Eq.(\ref{10}) are really observed in the experiment \cite{2} (these eigenstates of $K^0K^0$ were overlooked in the Ref.\cite{6}). \\
(b)\ The more interesting case occurs for $K^0\bar{K}^0$ pair created in the antisymmetric state with intensity given by Eq.(\ref{9}) being a function of $(t_a-t_b)$ (not $(t_a+t_b)$ as shown by Eq.(12) for symmetric states) which is proportional to $(t_1-t_2)$ in the $S$ system. In the EPR limit $t_1=t_2$, $K^0\bar{K}^0$ events dominate whereas like-strangeness events are strongly suppressed as shown by Eq.(\ref{8}) (see Fig.1 in \cite{2}). So the experimental facts remind us of the possibility that $K^0\bar{K}^0$ events may be related to common lowest (zero) eigenvalues of some commutative operators (just like what happened in Eqs.(\ref{14}) and (\ref{15}) for operators $\hat{A}$ and $\hat{B}$ (which are applied to symmetric states (due to $\hat{D}=x_1-x_2=v(t_1+t_2)$) but are not suitable for antisymmetric states), there are another three operators shown by Eqs.(\ref{4}) and (\ref{5}) being:
the operator of "flight-path difference" $\hat{F}=x_1+x_2=v(t_1-t_2)$, $\hat{M}=\hat{p}_1-\hat{p}_2$ and $\hat{G}=\hat{E}_1+\hat{E}_2$ with commutation relations as:
\begin{equation}\label{16}
[\hat{F},\hat{M}]=0,\;[\hat{F},\hat{G}]=0,\;[\hat{M},\hat{G}]=0
\end{equation}
which are just suitable for antisymmetric states. For $K^0\bar{K}^0$ back-to-back events, assume that one of two particles, say 2, is an antiparticle with its momentum and energy operators being
\begin{equation}\label{17}
\hat{p}_x^c=i\hbar\dfrac{\partial}{\partial x},\;\hat{E}^c=-i\hbar\dfrac{\partial}{\partial t}
\end{equation}
(the superscript $c$ means "antiparticle") just opposite in the sign to that for a particle. For instance, a freely moving particle's WF reads\footnotemark[1]\footnotetext[1]{Please see the derivation of Eqs.(2.19) and (2.20) from the quantum field theory (QFT) at Eqs.(5-7)-(5-9).}:
\begin{equation}\label{18}
\psi(x,t)\sim\exp\left[\frac{i}{\hbar}(px-Et)\right]
\end{equation}
whereas
\begin{equation}\label{19}
\psi_c(x,t)\sim\exp\left[-\frac{i}{\hbar}(p_cx-E_ct)\right]
\end{equation}
for its antiparticle with $p_c$ and $E_c\,(>0)$ being momentum and energy of the antiparticle in accordance with Eq.(\ref{17}). If using Eqs.(\ref{17})-(\ref{19}), we find
\begin{equation}\label{20}
\hat{F}\Psi^{antisym}_{K^0\bar{K}^0}(x_1,t_1;x_2,t_2)=F^{unlike}_k\Psi^{antisym}_{K^0\bar{K}^0}(x_1,t_1;x_2,t_2)=v(t_1-t_2)\Psi^{antisym}_{K^0\bar{K}^0}(x_1,t_1;x_2,t_2)
\end{equation}
with continuous index $k$ referring to continuous eigenvalues $F_k=v(t_1-t_2)$. Here, the WF in space-time of this system during measurement reads approximately:
\begin{equation}\label{21}
\Psi^{antisym}_{K^0\bar{K}^0}(x_1,t_1;x_2,t_2)\sim e^{i(p_1x_1-E_1t_1)}e^{-i(p_2^cx_2-E_2^ct_2)}
\end{equation}
with antiparticle 2 moving opposite to particle 1 and $p_2^c=-p_1$.

Now we use $\hat{M}(=\hat{p}_1-\hat{p}_2)=\hat{p}_1+\hat{p}^c_2$ on $K^0\bar{K}^0$ system, yielding
\begin{equation}\label{22}
\hat{M}\Psi^{antisym}_{K^0\bar{K}^0}(x_1,t_1;x_2,t_2)=M^{unlike}_k\Psi^{antisym}_{K^0\bar{K}^0}(x_1,t_1;x_2,t_2)
=(p_1+p^c_2)\Psi^{antisym}_{K^0\bar{K}^0}(x_1,t_1;x_2,t_2)
\end{equation}

Similarly, we have $\hat{G}(=\hat{E}_1+\hat{E}_2)=\hat{E}_1-\hat{E}^c_2$ and find
\begin{equation}\label{223}
\hat{G}\Psi^{antisym}_{K^0\bar{K}^0}(x_1,t_1;x_2,t_2)=G^{unlike}_k\Psi^{antisym}_{K^0\bar{K}^0}(x_1,t_1;x_2,t_2)
=(E_1-E^c_2)\Psi^{antisym}_{K^0\bar{K}^0}(x_1,t_1;x_2,t_2)
\end{equation}
Hence we see that once Eqs.(\ref{17}) and (\ref{19}) are accepted, the WFs $\Psi^{antisym}_{K^0\bar{K}^0}(x_1,t_1;x_2,t_2)$ show up in experiments as the only WFs with strongest intensity at the EPR limit ($t_1=t_2$) corresponding to their three eigenvalues being all zero: $F_k=M^{unlike}_k=G^{unlike}_k=0$ and they won't change even when accelerator's energies are going up.\\
If using Eq.(\ref{17}), the eigenvalues of $\hat{A}$ and $\hat{B}$ for the WF $\Psi^{antisym}_{K^0\bar{K}^0}(x_1,t_1;x_2,t_2)$ are $A^{unlike}_j=p_1-p^c_2=2p_1$ and $B^{unlike}_j=E_1+E^c_2=2E_1$ respectively, while that of $\hat{M}$ and $\hat{G}$ for the WF $\Psi^{antisym}_{K^0K^0}(x_1,t_1;x_2,t_2)$ are $M^{like}_k=p_1-p_2=2p_1$ and $G^{like}_k=E_1+E_2=2E_1$, respectively, those eigenvalues are much higher than zero and going up with the accelerator's energy.

Something is very interesting here: If we deny Eq.(\ref{17}) but insist on unified operators $\hat{p}$ and $\hat{E}$ for both particle and antiparticle, there would be no difference in eigenvalues between like-strangeness events and unlike-strangeness ones. For example, the $M^{unlike}_k$ and $G^{unlike}_k$ would be $2p_1$ and $2E_1$ too (instead of "0" as in Eqs.(\ref{22}) and (\ref{23})). This would mean that three commutative operators $\hat{F},\hat{M}$ and $\hat{G}$ are not enough to distinguish the WF $\Psi^{antisym}_{K^0\bar{K}^0}(x_1,t_1;x_2,t_2)$ from the WF $\Psi^{antisym}_{K^0K^0}(x_1,t_1;x_2,t_2)$ even they behave so differently as shown by Eqs.(\ref{8}) and (\ref{9})), especially at the EPR limit ($t_1=t_2$).\\
Eq.(\ref{17}) together with the identification of WF $\Psi^{antisym}_{K^0\bar{K}^0}(x_1,t_1;x_2,t_2)$ by three zero eigenvalues implies that the difference of a particle from its antiparticle is not something hiding in the "intrinsic space" like opposite charge (for electron and positron) or opposite strangeness (for $K^0$ and $\bar{K}^0$) but can be displayed in their WFs evolving in space-time at the level of QM. \\
To our knowledge, Eq.(\ref{17}) can be found at a page note of a paper by Konopinski and Mahmaud in 1953 \cite{9}, also appears in Refs.\cite{6,10,11,12,13}. \footnotemark[1]\footnotetext[1]{Eqs.(\ref{7})-(\ref{11}), the WFs and predicted intensities for $K^0-\bar{K}^0$ system, are derived from QFT (see Appendix I in Ref\cite{1}). On the other hand, our discussions about commutation relations, Eqs.(\ref{1})-(\ref{5}), and their eigenvalues of measurements, Eqs.(\ref{12})-(\ref{223}), are at the level of RQM, where two particles' WFs with their space-time coordinates are taken into account. Here we use Eqs.(\ref{7})-(\ref{11}) only to divide commutative operators into two groups, one for symmetric states, one for antisymmetric states. So the EPR problem is discussed at the level of RQM.}

\section{III. How to Make Klein-Gordon Equation a Self-Consistent Theory in RQM ?}
\label{sec:invariance}
\setcounter{equation}{0}
\renewcommand{\theequation}{3.\arabic{equation}}
\vskip 0.1cm

Let us begin with the energy conservation law for a particle in classical mechanics:
\begin{equation}\label{44}
E=\frac{1}{2m}{\bf p}^2+V({\bf x})
\end{equation}
Consider the rule promoting observables into operators:
\begin{equation}\label{45}
E\to\hat{E}=i\hbar\frac{\partial}{\partial t},\quad {\bf p}\to \hat{\bf p}=-i\hbar\nabla
\end{equation}
and let Eq.(\ref{44}) act on a wavefunction (WF) $\psi({\bf x},t)$, the Schr\"{o}dinger equation
\begin{equation}\label{46}
i\hbar\frac{\partial}{\partial t}\psi({\bf x},t)=-\frac{\hbar^2}{2m}\nabla^2\psi({\bf x},t)+V({\bf x})\psi({\bf x},t)   \end{equation}
follows immediately. In mid 1920's, considering the kinematical relation for a particle in the theory of special relativity (SR):
\begin{equation}\label{47}
(E-V)^2=c^2{\bf p}^2+m^2c^4
\end{equation}
and using Eq.(\ref{45}) again, the Klein-Gordon (KG) equation was established as:
\begin{equation}\label{48}
(i\hbar\frac{\partial}{\partial t}-V)^2\psi({\bf x},t)=-c^2\hbar^2\nabla^2\psi({\bf x},t)+m^2c^4\psi({\bf x},t)    \end{equation}
For a free KG particle, its plane-wave solution reads:
\begin{equation}\label{49}
\psi({\bf x},t)\sim \exp[\frac{i}{\hbar}({\bf p}\cdot{\bf x}-Et)]
\end{equation}
However, two difficulties arose:

(a) The energy $E$ in Eq.(\ref{49}) has two eigenvalues:
\begin{equation}\label{50}
E=\pm\sqrt{c^2{\bf p}^2+m^2c^4}
\end{equation}
What the "negative energy" means?

(b)The continuity equation is derived from Eq.(\ref{48}) as
\begin{equation}\label{51}
\frac{\partial\rho}{\partial t}+\nabla\cdot{\bf j}=0
\end{equation}
where
\begin{equation}\label{52}
\rho=\frac{i\hbar}{2mc^2}(\psi^*\frac{\partial}{\partial t}\psi-\psi\frac{\partial}{\partial t}\psi^*)-\frac{1}{mc^2}V\psi^*\psi
\end{equation}
and
\begin{equation}\label{53}
{\bf j}=\frac{i\hbar}{2m}(\psi\nabla\psi^*-\psi^*\nabla\psi)
\end{equation}
are the "probability density" and "probability current density" respectively. While the latter is the same as that derived from Eq.(\ref{46}), Eq.(\ref{52}) seems not positive definite and dramatically different from $\rho=\psi^*\psi$ in Eq.(\ref{46}).

In 1958, Feshbach and Villars \cite{25} recast Eq.(\ref{48}) into two coupled Schr\"{o}dinger-like equations as:\footnotemark[1]\footnotetext[1]{Interestingly, if ignoring the coupling between $\phi$ and $\chi$ and $V=0$ in Eq.(\ref{54}), they satisfy respectively the "two equations" written down by Schr\"{o}dinger in his 6th paper in 1926 (Annaten der Physik (4) V.81, 1926, p.104)when he invented NRQM in the form of wave mechanics.}
\begin{equation}\label{54}
\left\{
  \begin{array}{ll}
   \left(i\hbar\dfrac{\partial}{\partial t}-V\right)\phi=mc^2\phi-\dfrac{\hbar^2}{2m}\nabla^2(\phi+\chi)    \\[4mm]
    \left(i\hbar\dfrac{\partial}{\partial t}-V\right)\chi=-mc^2\chi+\dfrac{\hbar^2}{2m}\nabla^2(\phi+\chi)
  \end{array}
\right.
\end{equation}
where
\begin{equation}\label{55}
\left\{
  \begin{array}{ll}
   \phi=\dfrac{1}{2}\left[\left(1-\dfrac{1}{mc^2}V\right)\psi+\dfrac{i\hbar}{mc^2}\dot{\psi}\right]    \\[4mm]
    \chi=\dfrac{1}{2}\left[\left(1+\dfrac{1}{mc^2}V\right)\psi-\dfrac{i\hbar}{mc^2}\dot{\psi}\right]
  \end{array}
\right.
\end{equation}
($\dot{\psi}=\frac{\partial\psi}{\partial t}$). Interestingly, while $\psi=\phi+\chi$, the "probability density", Eq.(\ref{52}) can be recast into a difference between two positive-definite densities \cite{6,8}:
\begin{equation}\label{56}
\rho=\phi^*\phi-\chi^*\chi
\end{equation}
For simplicity, consider a free KG particle ($V=0$) with WF Eq.(\ref{49}). Then
\begin{equation}\label{3-14}
\left\{
  \begin{array}{ll}
    \phi=\dfrac{1}{2}\left(1+\dfrac{E}{mc^2}\right)\psi &  \\
    \chi=\dfrac{1}{2}\left(1-\dfrac{E}{mc^2}\right)\psi &
  \end{array}
\right.
\end{equation}
So $|\phi|>|\chi|$ and $\rho>0$ for $E>0$ case. But for a negative-energy ($E<0$) particle, we would have $\rho_{E<0}=|\phi|^2-|\chi|^2<0$. Thus we see that the difficulty of negative probability density is intimately related to that of negative-energy. The later difficulty is actually solved in the last section by regarding the negative-energy WF of a particle directly as the positive-energy WF of its antiparticle and introducing operators as
\begin{equation}\label{3-15}
 \hat{E}_c=-i\hbar\dfrac{\partial}{\partial t},\quad \hat{\bf p}_c=i\hbar\nabla
\end{equation}
when these two operators act on the antiparticle's WF
\begin{equation}\label{3-16}
 \psi_c({\bf x},t)\sim \exp\left[-\dfrac{i}{\hbar}({\bf p}_c\cdot{\bf x}-E_ct)\right],\quad (E_c>0)
\end{equation}
The antiparticle's energy $E_c(>0)$ and momentum ${\bf p}_c$ will be picked up. If substituting $\psi_c$, Eq.(\ref{3-16}), into Eq.(\ref{55}) to replace $\psi$, we obtain (after adding subscript "c" for antiparticle, $\psi_c=\phi_c+\chi_c$ and $V=0$ again)
\begin{equation}\label{3-17}
\left\{
  \begin{array}{ll}
    \phi_c=\dfrac{1}{2}\left(1-\dfrac{E_c}{mc^2}\right)\psi_c &  \\
    \chi_c=\dfrac{1}{2}\left(1+\dfrac{E_c}{mc^2}\right)\psi_c &
  \end{array}
\right.
\end{equation}
Now $|\phi_c|<|\chi_c|$. To set a positive-definite probability density $\rho_c$ for describing the antiparticle, we need
\begin{equation}\label{3-18}
\rho_c=|\chi_c|^2-|\phi_c|^2>0
\end{equation}
instead of Eq.(\ref{56}) for particle. However, if we directly add Eq.(\ref{3-18}) for antiparticle, this would not be a good theory. So we should have a unified prescription to get everything right for KG equation. Inspecting Eq.(\ref{54}) carefully, we do find that it is invariant under a (newly defined) space-time inversion (${\bf x}\to -{\bf x},t\to -t$), \ie, there is a hidden symmetry as follows:
\begin{equation}\label{3-19}
\left\{
\begin{array}{l}
 V({\bf x},t)\to -V({\bf x},t)=V_c({\bf x},t),\\
 \psi({\bf x},t)\to \psi(-{\bf x},-t)=\psi_c({\bf x},t),\\
 \phi({\bf x},t)\to \phi(-{\bf x},-t)=\chi_c({\bf x},t),\\
  \chi({\bf x},t)\to \chi(-{\bf x},-t)=\phi_c({\bf x},t)
\end{array}\right.
\end{equation}
For example, space-time inversion will bring Eq.(\ref{45}) for particle into Eq.(\ref{3-15}) for antiparticle. Meanwhile, performing Eq.(\ref{3-19}) on Eq.(\ref{55}), we find $
\chi_c(\phi_c)$ satisfying the same equation of $\chi(\phi)$ , Eq.(\ref{54}), and they read:
\begin{equation}\label{3-20}
\left\{
\begin{array}{l}
 \chi_c=\dfrac{1}{2}\left[\left(1+\dfrac{1}{mc^2}V\right)\psi_c-\dfrac{i\hbar}{mc^2}\dot{\psi}_c\right]\\[4mm]
\phi_c=\dfrac{1}{2}\left[\left(1-\dfrac{1}{mc^2}V\right)\psi_c+\dfrac{i\hbar}{mc^2}\dot{\psi}_c\right]
\end{array}\right.
\end{equation}
which is in conformity with the transformation of $\rho$, Eq.(\ref{52}), as expected:
\begin{equation}\label{3-21}
\rho\to\rho_c=\frac{i\hbar}{2mc^2}(\psi_c\dot{\psi}_c^*-\psi_c^*\dot{\psi}_c)+\frac{1}{mc^2}V\psi_c^*\psi_c
=\chi_c^*\chi_c - \phi_c^*\phi_c>0
\end{equation}
Similarly, we have
\begin{equation}\label{61}
{\bf j}\to{\bf j}_c=\frac{i\hbar}{2m}(\psi^*_c\nabla{\psi}_c-\psi_c\nabla\psi^*_c)
\end{equation}
which, for $V=0$ case, means
\begin{equation}\label{61}
{\bf j}=\dfrac{{\bf p}}{m}|\psi|^2\to{\bf j}_c=\dfrac{{\bf p}_c}{m}|\psi_c|^2,\;\;(V=0)
\end{equation}
with ${\bf j}_c$ along the direction of ${\bf p}_c$ as expected. Eq.(\ref{51}) remains valid after the above transformation too. Thus we see that both the "probability density" $\rho$ for a particle and $\rho_c$ for an antiparticle are positive definite before they can be normalized as expected:
\begin{equation}\label{62}
\int\rho d^3x=\int\rho_c d^3x=1
\end{equation}
Hence we see that the space-time inversion Eq.(\ref{3-19}) reflects the underlying symmetry between particle and antiparticle and overcomes two difficulties of KG equation simultaneously in an elegant way.

Here, we would like to introduce a "mass inversion" which can realize the same symmetry of Eq.(\ref{3-18})as follows:
\begin{equation}\label{3-25}
\left\{
\begin{array}{l}
m\to -m\\
 V({\bf x},t)\to V({\bf x},t),\\
 \psi({\bf x},t)\to \psi_c({\bf x},t),\\
 \phi({\bf x},t)\to \chi_c({\bf x},t),\\
  \chi({\bf x},t)\to \phi_c({\bf x},t)
\end{array}\right.
\end{equation}
Notice that, when $m\to-m$, we have $p\to -p_c$ and $E\to-E_c$ for a free particle transforming into its antiparticle because the momentum and energy in their WFs are proportional to the mass in SR. \footnotemark[1]\footnotetext[1]{Here $m$ always refers to the "rest mass" for a particle or its antiparticle, see the excellent paper by Okun in Ref.\cite{23}.}

The reason why $V\to -V$ in the space-time inversion Eq.(\ref{3-19}) whereas $V\to V$ in the mass inversion Eq.(\ref{3-25}) can be seen from the classical equation: The Lorentz force ${\bf F}$ on a particle exerted by an external potential $\Phi$ reads: ${\bf F}=-\nabla V=-\nabla(q\Phi)=m{\bf a}$. As the acceleration ${\bf a}$ of particle will change to $-{\bf a}$ for its antiparticle, there are two alternative explanations: either due to the inversion of charge $q\to-q$ (\ie, $V\to -V$ but keeping $m$ unchanged) or due to the inversion of mass $m\to-m$ (but keeping $V$ unchanged).

The reason why Feshbach-Villars' dissociation of KG equation, Eq.(\ref{54}), is so important is because they unveiled a new point of view for us to see a particle as follows:

For a free KG particle (say, $K^-$ meson) moving at a high speed $v$, its WF $\psi\sim e^{-iEt}\,(E>0)$ is always composed of two fields, $\phi$ and $\chi$, in confrontation as shown by Eqs.(\ref{54}) and (\ref{55}). Calculations (as can be
seen from Eq.(\ref{3-14})) show that: as long as $E>0$, then $|\phi|>|\chi|$ and $\rho>0$, so $\phi$ dominates $\chi$ and $K^-$ remains as a particle. However, the amplitude of $\chi$ increases with the increase of particle's energy $E$: when $v\to0, E\to m, |\chi|\to0$, but when $E\to\infty, |\chi|\to |\phi|$, the ratio between them reads: $|\chi|/|\phi|=
[1-(1-v^2/c^2)^{1/2}]/[1+(1-v^2/c^2)^{1/2}]$. What does this mean? It seems to us that while $\phi$ (hidden in $\psi$) characterizes the particle's property, $\chi$ represents the hidden "antiparticle (say, $K^+$) field" in the WF of this $K^-$ particle. Indeed, in Eqs.(\ref{49}) and (\ref{3-16}), the WFs of particle (dominated by $\phi$) and antiparticle (dominated by $\chi_c$), their phase variations with respect to space-time (keeping the same values of momentum and energy) are just in opposite directions, meaning that the intrinsic tendencies of space-time evolution of $\phi$ and $\chi$ are also in opposite directions essentially (see Eq.(\ref{54}) with $V=0$). Hence, even the $\chi$ is in a subordinate position in a particle, it still strives to display itself as follows: On one hand, $\chi$ holds $\phi$ back from going forward in space, so the particle's velocity $v$ has an upper limit value $c$ when $|\chi|$ approaches $|\phi|$. And during the "boosting" process of a particle's wave-packet, it shows the Lorentz contraction due to the entanglement between $\phi$ and $\chi$ (see calculation shown in Fig.9.5.1 of Ref.\cite{11}). On the other hand, a clock attached to the particle will show the time dilatation effect in SR with the increase of velocity $v$. This is because, in some sense, the "intrinsic clock" of $\phi(\chi)$ is running clockwise (anticlockwise). With the enhancement of $\chi$, the particle's clock, though still runs clockwise, tends to stop. Therefore, it seems to us that all SR effects of a particle could be calculated and understood by the existence and enhancement of "hidden antiparticle field" $\chi$ inside (for detail, please see \cite{11}).

Now we are going to prove that in RQM the antiparticle's WF $\psi_c$ obtained from the (newly defined) space-time inversion is coinciding with the CPT transformation of particle's WF $\psi$, but not that from $C\psi$.

As is well known, these three discrete transformations of $C,P$ and $T$ in RQM were defined for spinless particles separately as follows:

(a) Space-inversion ($P$):

The sign change of space coordinates (${\bf x}\to -{\bf x}$) in the wave function ($WF$) of $QM$ may lead to two eigenstates:
\begin{equation}\label{23}
\psi_{\pm}({\bf x},t)\to P\psi_{\pm}({\bf x},t)=\psi_{\pm}(-{\bf x},t)=\pm\psi_{\pm}({\bf x},t)
\end{equation}
with eigenvalues $1$ or $-1$ being the even or odd parity.

(b) Time reversal ($T$):

The so-called $T$ transformation is actually not a "time reversal" but a "reversal of motion" \cite{11,16}, which implies an antiunitary operator acting on the WF:
\begin{equation}\label{24}
\psi({\bf x},t)\to T\psi({\bf x},t)=\psi^*({\bf x},-t)
\end{equation}
(c) Charge conjugation transformation($C$):

The $C$ transformation brings a particle (with charge $q$) into its antiparticle (with charge $-q$) and implies
a complex conjugation on the $WF$:
\begin{equation}\label{25}
\psi({\bf x},t)\to C\psi({\bf x},t)= \psi^*({\bf x},t)
\end{equation}
Note that the $WF$ $\psi^*$ implies a negative-energy particle. Usually, one has to resort
to so-called "hole theory" for electron --- the vacuum is fully filled with infinite negative-energy electrons and a
"hole" created in the "sea" would correspond to a positron\cite{15,17}. But how could the "hole theory" be applied to
the boson particle? No one knows.

Fortunately, at the level of RQM, if one performs the product operator CPT on a particle's WF $\psi({\bf x},t)$, two operations of complex conjugation in the C and T will cancel each other, yielding \cite{7,17}
\begin{equation}\label{3-29}
 \psi({\bf x},t)\to CPT\psi({\bf x},t)=\psi_{CPT}({\bf x},t)=\psi(-{\bf x},-t)
\end{equation}
On the right-side-hand (RHS), the $\psi_{CPT}({\bf x},t)$ is just the antiparticle's WF obtained from the (newly defined) space-time inversion (${\bf x}\to -{\bf x},t\to-t$), Eq.(\ref{3-16}).

One may ask: Is the $C\psi=\psi^*$ in Eq.(\ref{25}) gives the same space-time evolution behavior of EQ.(\ref{3-29}) or Eq.(\ref{3-16})? The answer is "yes" and "no". We say "yes" because it seems formally correct but "no" because it is essentially incorrect --- given one set of operators like Eq.(\ref{45}) for both $\psi$ and $C\psi=\psi^*$, then the $C\psi$ is a negative energy WF, which cannot be accepted in physics. Moreover, as we will see in the next section, for Dirac equation, while the particle's WF $\psi({\bf x},t)$ and its antiparticle's $\psi_c({\bf x},t)=\psi_{CPT}({\bf x},t)\sim \psi(-{\bf x},-t)$ describes the same energy and momentum, they must have opposite helicities. Given the definition of C, $C\psi=\psi^*$ is bound to fail in this aspect too.

Above discussions are strictly at the level of RQM. We have shown that the KG equation is really a  self-sufficient and simplest model in RQM as long as the (newly defined) space-time inversion is introduced. As we will continue to do so in next sections, let us write down it in a compact equation as a renamed CPT invariance:
\begin{equation}\label{3-30}
{\cal P}{\cal T}={\cal C}
\end{equation}
Here, ${\cal P}{\cal T}$ represent (${\bf x}\to-{\bf x},t\to-t$), \ie, the "strong reflection" invented by L\"{u}ders and Pauli (1954-1957) in their proof of CPT theorem at the level of QFT, and ${\cal C}$, the newly defined particle-antiparticle transformation operator, whose definition is precisely contained in Eq.(\ref{3-30}).

In 1965, Lee and Wu proposed that\cite{21}:
\begin{equation}\label{3-31}
 |\bar{a}\rangle=CPT|a\rangle
\end{equation}
where $|a\rangle$ and $|\bar{a}\rangle$ represent a particle and its antiparticle. To our understanding, the physical essence of Eq. (\ref{3-31}) is just that claimed by Eq. (\ref{3-30}). What we add in Eq.(\ref{3-30}) here is: It can be derived at the level of RQM much easier than that at the level of QFT, as long as we admit that "performing the operation of complex conjugation twice means doing nothing at all", so what left after CPT product operation in RQM is merely the space-time inversion operation, ${\bf x}\to-{\bf x},t\to-t$. And we should accept Eq.(\ref{3-15}) for antiparticle (so $\psi(-{\bf x},-t)=\psi_c({\bf x},t)$) and the underlying symmetry Eq.(\ref{3-19}) to reach Eq.(\ref{3-30}). The ${\cal C}$ operator in Eq.(\ref{3-30}) shows that a particle's "intrinsic property" is intimately linked to the space-time.

Based on the ideas and method in this section, the Klein paradox for KG equation can be solved as discussed in the Appendix A.

\section{IV. Dirac Equation as Coupled equations of two-component Spinors}
\label{sec:DiracEquation}
\setcounter{equation}{0}
\renewcommand{\theequation}{4.\arabic{equation}}
\vskip 0.1cm

Let us turn to the Dirac equation describing an electron
\begin{equation}\label{66}
\left(i\hbar\dfrac{\partial}{\partial t}-V\right)\psi=H\psi=(-i\hbar c{\boldsymbol\alpha}\cdot\nabla+\beta mc^2)\psi
\end{equation}
with ${\boldsymbol\alpha}$ and $\beta$ being $4\times4$ matrices, the WF $\psi$ is a four-component spinor
\begin{equation}\label{67}
\psi=\begin{pmatrix}\phi\\ \chi\end{pmatrix}
\end{equation}
Usually, the two-component spinors $\phi$ and $\chi$ are called "positive" and "negative" energy components. In our point of view, they are the hiding "particle" and "antiparticle" fields in a particle (electron) respectively (\cite{11}, see below). Substitution of Eq.(\ref{67}) into Eq.(\ref{66}) leads to
\begin{equation}\label{68}
\left\{
\begin{array}{l}
\left(i\hbar\dfrac{\partial}{\partial t}-V\right)\phi=-i\hbar c{\boldsymbol\sigma}\cdot\nabla\chi+mc^2\phi\\[3mm]
\left(i\hbar\dfrac{\partial}{\partial t}-V\right)\chi=-i\hbar c{\boldsymbol\sigma}\cdot\nabla\phi-mc^2\chi
\end{array}\right.
\end{equation}
(${\boldsymbol\sigma}$ are Pauli matrices). Eq.(\ref{68}) is invariant under the space-time inversion $({\bf x}\to -{\bf x},t\to -t)$ with
\begin{equation}\label{69}
\left\{
\begin{array}{l}
\phi({\bf x},t)\to \phi(-{\bf x},-t)=\chi_c({\bf x},t),\;\chi({\bf x},t)\to \chi(-{\bf x},-t)=\phi_c({\bf x},t)\\
V({\bf x},t)\to -V({\bf x},t)=V_c({\bf x},t)
\end{array}\right.
\end{equation}
again, showing that Dirac equation is in conformity with the underlying symmetry Eq.(\ref{3-30}).
Note that under the space-time inversion, the ${\boldsymbol\sigma}$ remain unchanged (However, see  Eqs.(\ref{74})-(\ref{76}) below). Alternatively, Eq.(\ref{68}) also remains invariant under a mass inversion as
\begin{equation}\label{70}
m\to -m,\;\phi({\bf x},t)\to\chi_c({\bf x},t),\;\chi({\bf x},t)\to\phi_c({\bf x},t),\; V\to V
\end{equation}
In either case of Eq.(\ref{69}) or (\ref{70}), we have\footnotemark[2]\footnotetext[2]{The reason why we use $\psi'_c$ instead of $\psi_c$ will be clear in Eqs.(\ref{77})-(\ref{80}). Actually, we emphasize Dirac equation as a coupling equation of two two-component spinors, Eq.(\ref{68}), rather than merely a four-component spinor equation.}
\begin{equation}\label{71}
\psi({\bf x},t)=\begin{pmatrix}\phi({\bf x},t)\\ \chi({\bf x},t)\end{pmatrix}\to \begin{pmatrix}\chi_c({\bf x},t)\\ \phi_c({\bf x},t)\end{pmatrix}=\psi'_c({\bf x},t)
\end{equation}
For concreteness, we consider a free electron moving along the $z$ axis with momentum $p=p_z>0$ and having a helicity $h={\boldsymbol\sigma}\cdot{\bf p}/|{\bf p}|=1$, its WF reads:
\begin{equation}\label{72}
\psi(z,t)\sim \begin{pmatrix}\phi\\ \chi\end{pmatrix}\sim
\begin{pmatrix}1\\ 0\\ \frac{p}{E+m}\\ 0\end{pmatrix}\exp[i(pz-Et)]
\end{equation}
with $|\phi|>|\chi|$. Under a space-time inversion ($z\to -z,t\to -t,p\to p_c,E\to E_c$) or mass inversion ($m\to -m,p\to -p_c,E\to -E_c$), it transforms into a WF for positron (moving along $z$ axis)
\begin{equation}\label{73}
\psi'_c(z,t)\sim \begin{pmatrix}\chi_c\\ \phi_c\end{pmatrix}\sim
\begin{pmatrix}1\\ 0\\ \frac{p_c}{E_c+m}\\ 0\end{pmatrix}\exp[-i(p_cz-E_ct)]
\end{equation}
with $|\chi_c|>|\phi_c|,\;(p_c>0,E_c>0)$. However, the positron's helicity becomes $h_c={\boldsymbol\sigma}_c\cdot{\bf p}_c/|{\bf p}_c|=-1$. This is because the total angular momentum operator for an electron reads
\begin{equation}\label{74}
\hat{\bf J}=\hat{\bf L}+\frac{\hbar}{2}{\boldsymbol\sigma}
\end{equation}
Under a space-time inversion, the orbital angular momentum operator transforms as
\begin{equation}\label{75}
\hat{\bf L}={\bf r}\times\hat{\bf p}={\bf r}\times(-i\hbar\nabla)\to{-\bf r}\times(i\hbar\nabla)
=-{\bf r}\times\hat{\bf p}_c=-\hat{\bf L}_c
\end{equation}
To get $\hat{\bf j}\to -\hat{\bf j}_c$ with $\hat{\bf j}_c=\hat{L}_c+\frac{\hbar}{2}\hat{\boldsymbol\sigma}_c$, we should have
\begin{equation}\label{76}
\hat{\boldsymbol\sigma}_c=-\hat{\boldsymbol\sigma}
\end{equation}
Hence the values of matrix element for positron's spin operator ${\boldsymbol\sigma}_c$ is just the negative to that for ${\boldsymbol\sigma}$ in the same matrix representation.

Notice that Eq.(\ref{72}) describes an electron with positive helicity, \ie, ${\boldsymbol\Sigma}\cdot\hat{{\bf p}}\psi=p_z\psi=p\psi$ \footnotemark[2]\footnotetext[2]{${\boldsymbol\Sigma}=\begin{pmatrix}{\boldsymbol\sigma}&0\\ 0&{\boldsymbol\sigma}\end{pmatrix},\;{\boldsymbol\Sigma}_c=\begin{pmatrix}{\boldsymbol\sigma}_c&0\\ 0&{\boldsymbol\sigma}_c\end{pmatrix}$}. Under a space-time inversion, it transforms into $(-{\boldsymbol\Sigma}_c)\cdot\hat{{\bf p}_c}\psi'_c=\Sigma_z(i\hbar\frac{\partial}{\partial z})\psi'_c=p_c\psi'_c$ in Eq.(\ref{73}), \ie, ${\boldsymbol\Sigma}_c\cdot\hat{{\bf p}_c}\psi'_c=-p_c\psi'_c$, meaning that Eq.(\ref{73}) describes a positron with negative helicity.

Dirac equation is usually written in a covariant form as (Pauli metric is used: $x_4=ict, \gamma_k=-i\beta\alpha_k, \gamma_4=\beta, \gamma_5=\gamma_1\gamma_2\gamma_3\gamma_4=-\begin{pmatrix}0 & I\\ I & 0\end{pmatrix}$, see \cite{15}):
\begin{equation}\label{77}
 (\gamma_\mu\partial_\mu+m)\psi=0
\end{equation}
Under a space-time (or mass) inversion, it turns into an equation for antiparticle:
\begin{equation}\label{78}
 (-\gamma_\mu\partial_\mu+m)\psi'_c=0
\end{equation}
with an example of $\psi'_c$ shown in Eq.(\ref{73}). Let us perform a representation transformation:
\begin{equation}\label{79}
\psi'_c\to\psi_c=(-\gamma_5)\psi'_c=\begin{pmatrix}\phi_c\\ \chi_c\end{pmatrix}
\end{equation}
and arrive at
\begin{equation}\label{80}
 (\gamma_\mu\partial_\mu+m)\psi_c=0
\end{equation}
due to $\{\gamma_5,\gamma_\mu\}=0$. Since $\psi_c$ and $\psi'_c$ are essentially the same in physics, (this is obviously seen from its resolved form, Eq.(\ref{68})), it is merely a trivial thing to change the position of $\chi_c$ in the 4-component spinor (lower in Eq.(\ref{79}) and upper in Eq.(\ref{73})). What important is $|\chi_c|>|\phi_c|$ for characterizing an  antiparticle versus $|\phi|>|\chi|$ for a particle. Therefore, if a particle with energy $E$ runs into a potential barrier $V=V_0>E+m$, its kinetic energy becomes negative, and its WF's third component in Eq.(\ref{72}) suddenly turns into $\frac{p'}{E-V_0+m}=\frac{-p'}{V_0-E-m},(p'=\sqrt{(E-V_0)^2-m^2})$, whose absolute magnitude is larger than that of the first component. This means that it is an antiparticle's WF satisfying Eq.(\ref{80}) (with $E_c=V_0-E(>m)$ and  $|\chi_c|>|\phi_c|$) and will be crucial for the explanation of Klein paradox in Dirac equation as shown in the Appendix A. However, we need to discuss the "probability density" $\rho$ and "probability current density" ${\bf j}$ for a Dirac particle versus $\rho_c$ and ${\bf j}_c$ for its antiparticle. Different from that in KG equation, now we have
\begin{equation}\label{81}
\rho=\psi^\dag\psi=\phi^\dag\phi+\chi^\dag\chi\to\rho_c=\psi_c^\dag\psi_c=\chi_c^\dag\chi_c+\phi_c^\dag\phi_c
\end{equation}
which is positive definite for either particle or antiparticle. On the other hand, we have
\begin{equation}\label{82}
{\bf j}=c\psi^\dag{\boldsymbol\alpha}\psi=c(\phi^\dag{\boldsymbol\sigma}\chi+\chi^\dag{\boldsymbol\sigma}\phi)
\to{\bf j}_c=c\psi_c^\dag{\boldsymbol\alpha}\psi_c=c(\chi_c^\dag{\boldsymbol\sigma}\phi_c+\phi_c^\dag{\boldsymbol\sigma}\chi_c)
\end{equation}
(we prefer to keep ${\boldsymbol\sigma}$ rather than ${\boldsymbol\sigma}_c$ for antiparticle). For Eqs.(\ref{72}), (\ref{73}) and (\ref{79}), we find ($c=\hbar=1$)
\begin{equation}\label{83}
j_z\sim\dfrac{2p}{E+m}>0\to j_z^c\sim\dfrac{2p_c}{E_c+m}>0\quad (V=0)
\end{equation}
which means that the probability current is always along the momentum's direction for either a particle or antiparticle.

Above discussions at RQM level may be summarized as follows: The first symptom for the appearance of an antiparticle is: If we perform an energy operator ( $E=i\hbar \partial / \partial t$) on a WF and find a negative energy ($E<0$) or a negative kinetic energy ($E-V <0$), we'd better to doubt the WF being a description of antiparticle and use the operators for antiparticle, Eq.(\ref{3-15}). Then for further confirmation, two more criterions for $\rho$ and ${\bf j}$ are needed (see Appendix).

In hindsight, for a linear equation in RQM, either KG or Dirac equation, the emergence of both positive and negative energy ($E$) WFs is inevitable and natural. From mathematical point of view, the set of WFs cannot be complete if without taking the negative energy solutions into account. And physicists believe that these negative-energy solutions might be relevant to antiparticles. However, we physicists admit that both a rest particle's energy $E=mc^2$ and a rest antiparticle's energy $E_c=m_cc^2=mc^2$ are positive, as verified by the experiments of pair-creation process $\gamma\to e^++e^-$. The above contradiction constructs so-called "negative-energy paradox" in RQM. For Dirac particle, majority (not all) of physicists accept the "hole theory" to explain the "paradox". But for KG particle, no such kind of "hole theory" can be acceptable. It was this "negative-energy paradox" and "Klein paradox" as well as the four "commutation relations", Eqs.(\ref{1})-(\ref{5}), hidden in the two-particle system discussed by EPR \cite{1} gradually prompted us to realize that the root cause of difficulty in RQM lies in an a priori notion --- only one kind of WF with one set of operators (like Eq.(\ref{45})) can be acceptable in QM, either for NRQM or RQM.

Once getting rid of the constraint in the above notion and introducing two sets of WFs and operators for particle and antiparticle respectively, we are able to see that many difficulties in RQM disappear immediately. What we emphasize in section III and IV is: The CPT invariance in RQM, \ie, the invariance of (newly defined) space-time inversion Eq.(\ref{3-30}) (which dictates Eq.(\ref{3-15}), $\rho_c$ and ${\bf j}_c$ for antiparticle) is capable of helping the RQM to become complete and more useful in applications (see Appendix A).

\section{V. Why QFT is Correct from Scratch?}
\label{sec:qft}
\setcounter{equation}{0}
\renewcommand{\theequation}{5.\arabic{equation}}
\vskip 0.1cm

In QFT, L\"{u}ders and Pauli proved the CPT theorem (Refs.\cite{31,32,33,34}), claiming that "a wide class of QFTs which are invariant under the proper Lorentz group is also invariant with respect to the product of $T,C$ and $P$". The proof of CPT theorem contains a crucial step being the construction of so-called "strong reflection", consisting in a reflection of space and time about some arbitrarily chosen origin, \ie, ${\bf r}\to -{\bf r},t\to -t$.

Pauli first proposed and explained the strong reflection in \cite{33} as follows: When the space-time coordinates change their sign, every particle transforms into its antiparticle simultaneously. The physical sense of the strong reflection is the substitution of every emission (absorption) operator of a particle by the corresponding absorption (emission) operator of its antiparticle. And there is no need to reverse the sign of the electric charge when the sign of space-time coordinates is reversed.

After combining with other necessary postulates, L\"{u}ders and Pauli proved that the Hamiltonian density of QFT (constructed from fields of spin zero, one-half and one by local interactions which are invariant under the proper Lorentz group) is invariant with respect to the strong reflection,  ${\cal H}({\bf x},t)\to {\cal H}(-{\bf x},-t)$. The strong reflection, after combining with the Hermitian conjugation (H.C.), turns to be identical with the product of $T, C$ and $P$ operations in QFT, thus completing the proof of CPT theorem \cite{32}.

Hence we can rename the CPT invariance and write down Eq.(\ref{3-30}) again, but at the level of QFT:
\begin{equation}\label{5-1}
{\cal P}{\cal T}={\cal C}
\end{equation}
It's time to look at the CPT theorem upside down as follows:

For a theorem, either in mathematics or in physics, its consequences are already hidden in its premises (which are essentially beyond the proof of theorem itself). Evidently, the premise of CPT theorem is QFT, but the premise of QFT had not been unveiled explicitly until 1954-1957. The great merit of L\"{u}ders and Pauli is: they discovered the premise of QFT is just the invariance of the "strong reflection"(in combination with the H.C.), or equivalently, the CPT invariance. In other words, why CPT theorem looks so unique is just because its consequence proves its premise exactly.

We are encouraged to say so because the validity of CPT invariance, \ie, the "strong reflection", has also been proved at the level of RQM in last two sections. Below, we highlight Pauli-L\"{u}ders' idea to show that the "field operator" in QFT is defined precisely in accordance with the CPT invariance, \ie, Eq.(\ref{5-1}). What we need is some supplement added in the Fock space according to Pauli's idea:

Let us look at the free "field operator" of charged scalar boson field, \ie, the complex KG field and its hermitian conjugate being defined as ($\hbar=c=1$)
\begin{equation}\label{5-2}
\left\{
  \begin{array}{ll}
   \hat{\psi}({\bf x},t)=\sum\limits_{\bf p}\dfrac{1}{\sqrt{2V\omega_{\bf p}}}\left\{\hat{a}_{\bf p}\exp[i({\bf p}\cdot{\bf x}-Et)]
+\hat{b}^\dag_{\bf p}\exp[-i({\bf p}\cdot{\bf x}-Et)]\right\}\\[4mm]
   \hat{\psi}^\dag({\bf x},t)=\sum\limits_{\bf p}\dfrac{1}{\sqrt{2V\omega_{\bf p}}}\left\{\hat{a}^\dag_{\bf p}\exp[-i({\bf p}\cdot{\bf x}-Et)]
+\hat{b}_{\bf p}\exp[i({\bf p}\cdot{\bf x}-Et)]\right\}
  \end{array}
\right.
\end{equation}
As Eq.(\ref{5-1}) works at the level of QFT, we expect that Eq.(\ref{5-2}) remains invariant under the operation ${\cal P}{\cal T}={\cal C}$ in the sense of
\begin{equation}\label{5-3}\begin{array}{l}
 \hat{\psi}({\bf x},t)\to{\cal P}{\cal T}\hat{\psi}({\bf x},t)({\cal P}{\cal T})^{-1}=\hat{\psi}(-{\bf x},-t)=\hat{\psi}({\bf x},t),\\
\hat{\psi}^\dag({\bf x},t)\to{\cal P}{\cal T}\hat{\psi}^\dag({\bf x},t)({\cal P}{\cal T})^{-1}=\hat{\psi}^\dag(-{\bf x},-t)=\hat{\psi}^\dag({\bf x},t)
\end{array}\end{equation}
Indeed, it does, as long as a transformation of operators in Fock space is expressed by \cite{11,28}:
\begin{equation}\label{5-4}
 \hat{a}_{\bf p}\leftrightarrows \hat{b}^\dag_{\bf p}
\end{equation}
simultaneously together with the inversion of space-time coordinates (${\bf x}\to -{\bf x},t\to-t$) in c-numbers WFs in Eq.(\ref{5-2}). Eqs.(\ref{5-2})-(\ref{5-4}) imply that under the space-time inversion, the process of a particle's annihilation transformations into that of its antiparticle's creation (or vice versa), an ansatz could be understood as a necessary implementation of Eq.(\ref{5-1}) at the level of QFT and further reflects Pauli's idea that the space-time inversion is indivisible from the transformation between particle and antiparticle. \footnotemark[1]\footnotetext[1]{Unlike in RQM, the "mass inversion" is not applicable in QFT.}

The commutation relation between $\hat{a}_{\bf p}$ and $\hat{a}^\dag_{\bf p}$ is assumed as usual:
\begin{equation}\label{5-5}
[\hat{a}_{\bf p},\hat{a}^\dag_{\bf p'}]=\delta_{\bf pp'},\quad [\hat{a}_{\bf p},\hat{b}_{\bf p'}]=0
\end{equation}
Then, performing Eq.(\ref{5-1}) on Eq.(\ref{5-5}), we arrive at:
\begin{equation}\label{5-6}
[\hat{b}_{\bf p},\hat{b}^\dag_{\bf p'}]=\delta_{\bf pp'},\quad [\hat{a}^\dag_{\bf p},\hat{b}^\dag_{\bf p'}]=0
\end{equation}
where Eq.(\ref{5-4}) has been used and one more rule is added as follows\cite{32}: The order of an operator product in Fock space has to be reversed under the space-time inversion, \eg, $({\cal P}{\cal T})\hat{A}\hat{B}({\cal P}{\cal T})^{-1}=({\cal P}{\cal T})\hat{B}({\cal P}{\cal T})^{-1}({\cal P}{\cal T})\hat{A}({\cal P}{\cal T})^{-1}$. So is the order of a process occurred in a many-particle system under the operation Eq.(\ref{5-1}). This is a necessary postulate (together with the Hermiticity of the Hamiltonian density) for QFT being capable of dealing with real problems successfully.

As QFT is such a successful theory, we may expect that the WFs for a particle and its antiparticle derived from the QFT will also be identified with that in the RQM.

In QFT, the WF of a single particle should be defined rigorously as the nondiagonal matrix element of the relevant field operator between the vacuum state and one-particle state. For instance, assume Eq.(\ref{5-2}) to be the "field operator of $K^-$ meson field", then the WF of a freely moving $K^-$ meson (with momentum ${\bf p}_1$) is given by
\begin{equation}\label{5-7}
\psi_{K^-}({\bf x},t)=\langle0|\hat{\psi}({\bf x},t)|K^-,{\bf p}_1\rangle=\langle0|\hat{\psi}({\bf x},t)\hat{a}^\dag_{{\bf p}_1}|0\rangle
=\dfrac{1}{\sqrt{2V\omega_{{\bf p}_1}}}e^{i({\bf p}_1\cdot{\bf x}-E_1t)}
\end{equation}
whereas the hermitian (\ie, complex) conjugation of a $K^+$ meson's WF is given by
\begin{equation}\label{5-8}
\psi^*_{K^+}({\bf x},t)=\langle0|\hat{\psi}^\dag({\bf x},t)|K^+,{\bf p}_1\rangle=\langle0|\hat{\psi}^\dag({\bf x},t)\hat{b}^\dag_{{\bf p}_1}|0\rangle
=\dfrac{1}{\sqrt{2V\omega_{{\bf p}_1}}}e^{i({\bf p}_1\cdot{\bf x}-E_1t)}
\end{equation}
which leads to $K^+$'s WF being
\begin{equation}\label{5-9}
\psi_{K^+}({\bf x},t)=\dfrac{1}{\sqrt{2V\omega_{{\bf p}_1}}}e^{-i({\bf p}_1\cdot{\bf x}-E_1t)}
\end{equation}
as expected.

Similarly, the "field operator" for Dirac field is constructed in the following form [see \cite{14,15}. However, instead of spin $s$ (projection along $z$ axis in space) usually used, here $h$, the helicity, is used in the expansion]:
\begin{equation}\label{5-10}
\left\{
  \begin{array}{ll}
   \hat{\psi}({\bf x},t)=\dfrac{1}{\sqrt{V}}\sum\limits_{\bf p}\sum\limits_{h=\pm1}\sqrt{\dfrac{m}{E}}\left[\hat{a}_{\bf p}^{(h)}u^{(h)}({\bf p})e^{i({\bf p}\cdot{\bf x}-Et)}+\hat{b}^{(h)\dag}_{\bf p}v^{(h)}({\bf p})e^{-i({\bf p}\cdot{\bf x}-Et)}\right]\\[4mm]
   \hat{\psi}^\dag({\bf x},t)=\dfrac{1}{\sqrt{V}}\sum\limits_{\bf p}\sum\limits_{h=\pm1}\sqrt{\dfrac{m}{E}}\left[\hat{a}_{\bf p}^{(h)\dag}u^{(h)\dag}({\bf p})e^{-i({\bf p}\cdot{\bf x}-Et)}+\hat{b}^{(h)}_{\bf p}v^{(h)\dag}({\bf p})e^{i({\bf p}\cdot{\bf x}-Et)}\right]
  \end{array}
\right.
\end{equation}
Here for arbitrary momentum ${\bf p}$ (with direction denoted by angles $\theta$ and $\phi$ in spherical coordinates) and energy $E=\sqrt{{\bf p}^2+m^2}>0$, the spinors attached to particle's annihilation operators $\hat{a}_{\bf p}^{(h)}$ are
\begin{equation}\label{5-11}
u^{(1)}({\bf p})=\sqrt{\dfrac{E+m}{2m}}\begin{pmatrix}\phi_0^{(1)}({\bf p})\\ \frac{|{\bf p}|}{E+m}\phi_0^{(1)}({\bf p})\end{pmatrix}
,\quad u^{(-1)}({\bf p})=\sqrt{\dfrac{E+m}{2m}}\begin{pmatrix}\phi_0^{(-1)}({\bf p})\\ \frac{-|{\bf p}|}{E+m}\phi_0^{(-1)}({\bf p})\end{pmatrix}
\end{equation}
\begin{equation}\label{5-12}
\phi_0^{(1)}({\bf p})=\begin{pmatrix}\cos\theta/2\\ e^{i\phi}\sin\theta/2\end{pmatrix},\;
\phi_0^{(-1)}({\bf p})=\begin{pmatrix}\sin\theta/2\\ -e^{i\phi}\cos\theta/2\end{pmatrix},\;
u^{(h)\dag}({\bf p})u^{(h')}({\bf p})=\dfrac{E}{m}\delta_{hh'}
\end{equation}
\begin{equation}\label{513}
\phi_0^{(h)}(-{\bf p})=\phi_0^{(-h)}({\bf p},\quad \gamma_4u^{(h)}(-{\bf p})=u^{(-h)}({\bf p})
\end{equation}
while that attached to antiparticle's creation operators $\hat{b}_{\bf p}^{(h)\dag}$ are
\begin{equation}\label{5-13}
v^{(h)}({\bf p})=(-\gamma_5)u^{(-h)}({\bf p}),\quad \gamma_4v^{(h)}(-{\bf p})=-v^{(-h)}({\bf p})
\end{equation}
\begin{equation}\label{5-14}
v^{(h)\dag}({\bf p})v^{(h')}({\bf p})=\dfrac{E}{m}\delta_{hh'},\;v^{(h')\dag}(-{\bf p})u^{(h)}({\bf p})
=u^{(h')\dag}(-{\bf p})v^{(h)}({\bf p})=0
\end{equation}
Like Eq.(\ref{5-4}) for KG field, an ansatz is added:
\begin{equation}\label{516}
\hat{a}_{\bf p}^{(h)}\rightleftarrows \hat{b}_{\bf p}^{(-h)\dag},\quad \hat{a}_{\bf p}^{(h)\dag}\rightleftarrows \hat{b}_{\bf p}^{(-h)}\quad ({\bf x}\to-{\bf x},t\to -t)
\end{equation}
with $h\rightleftarrows(-h)$ under the space-time inversion in complying with Eqs.(\ref{72})-(\ref{76}) and the discussion after them.

Different from that for KG field, now the operators $\hat{a}_{\bf p}^{(h)}$ and $\hat{b}_{\bf p}^{(h)}$ \etc are assumed to obey anticommutation relations:
\begin{equation}\label{5-16}
\{\hat{a}^{(h)}_{\bf p},\hat{a}^{(h')\dag}_{{\bf p}'}\}=\{\hat{b}^{(h)}_{\bf p},\hat{b}^{(h')\dag}_{{\bf p}'}\}=\delta_{{\bf p}{\bf p}'}\delta_{hh'},
\{\hat{a}^{(h)}_{\bf p},\hat{b}^{(h')}_{{\bf p}'}\}=0,\;\etc
\end{equation}
which, like Eqs.(\ref{5-5})-(\ref{5-6}) for KG field, remain invariant under the operation of Eq.(\ref{5-1}).

However, for Dirac field operator, we should define:
\begin{equation}\label{518}
\begin{array}{l}
\hat{\psi}({\bf x},t)\to {\cal P}{\cal T}\hat{\psi}({\bf x},t)({\cal P}{\cal T})^{-1}=-\gamma_5\hat{\psi}(-{\bf x},-t)=\hat{\psi}({\bf x},t)\\[4mm]
\hat{\psi}^\dag({\bf x},t)\to {\cal P}{\cal T}\hat{\psi}^\dag({\bf x},t)({\cal P}{\cal T})^{-1}=\hat{\psi}^\dag(-{\bf x},-t)(-\gamma_5)=\hat{\psi}^\dag({\bf x},t)
\end{array}
\end{equation}
for keeping their invariance in the 4-component spinor form rigorously. Thus
\begin{equation}\label{519}
\hat{\psi}(-{\bf x},-t)=-\gamma_5\hat{\psi}({\bf x},t),\quad \hat{\psi}^\dag(-{\bf x},-t)=\hat{\psi}^\dag({\bf x},t)(-\gamma_5)
\end{equation}
which are useful for proving the "spin-statistics connection" by ${\cal P}{\cal T}$ invariance.

Another rule is: One should always take the normal ordering when dealing with quadratic forms like $\hat{\bar\psi}(x)\psi(x)$ \etc

Now, like Eqs.(\ref{5-7})-(\ref{5-9}) for KG particles' WFs, for Dirac field, we have the WF of an electron being
\begin{equation}\label{5-17}
\psi_{e^-}({\bf x},t)=\langle0|\hat{\psi}({\bf x},t)|e^-,{\bf p}_1,h_1\rangle=\langle0|\hat{\psi}({\bf x},t)\hat{a}_{\bf p_1}^{(h_1)\dag}|0\rangle=\dfrac{1}{\sqrt V}\sqrt{\dfrac{m}{E_1}}u^{(h_1)}({\bf p}_1)e^{i({\bf p}_1\cdot{\bf x}-E_1t)}
\end{equation}
but the hermitian conjugate of a positron's WF is given by
\begin{equation}\label{5-18}
\psi_{e^+}^\dag({\bf x},t)=\langle0|\hat{\psi}^\dag({\bf x},t)|e^+,{\bf p}_c,h_c\rangle=\langle0|\hat{\psi}^\dag({\bf x},t)\hat{b}_{\bf p_c}^{(h_c)\dag}|0\rangle=\dfrac{1}{\sqrt V}\sqrt{\dfrac{m}{E_c}}v^{(h_c)\dag}({\bf p}_c)e^{i({\bf p}_c\cdot{\bf x}-E_ct)}
\end{equation}
which leads to positron's WF being
\begin{equation}\label{5-19}
\psi_{e^+}({\bf x},t)=\dfrac{1}{\sqrt V}\sqrt{\dfrac{m}{E_c}}v^{(h_c)}({\bf p}_c)e^{-i({\bf p}_c\cdot{\bf x}-E_ct)}
\end{equation}
as expected. To our surprise, we couldn't find such simple derivations like Eqs.(\ref{5-7})-(\ref{5-9}) and Eqs.(\ref{5-17})-(\ref{5-19}) in existing textbooks. \footnotemark[1]\footnotetext[1]{Notice that in Eq.(\ref{5-18}), the existence of spinor $v^{(h_c)\dag}({\bf p}_c)$ makes the claim for "hermitian conjugate" an indisputable one.}

Hence the historical merit of Pauli and L\"{u}ders could be highlighted as follows: On one hand, what they did is actually to correct a systematic error ("only one set of WF and operators is allowed") in RQM via the approach of QFT. On the other hand, they unveiled the underlying symmetry of QFT being the invariance of the Hamiltonian density ${\hat{\cal H}}({\bf x},t)$  under an operation of "strong reflection", \ie,
\begin{equation}\label{5-23}
{\hat{\cal H}}({\bf x},t)\to{\cal P}{\cal T}{\hat{\cal H}}({\bf x},t)({\cal P}{\cal T})^{-1}={\hat{\cal H}}(-{\bf x},-t)={\hat{\cal H}}({\bf x},t)
\end{equation}
as well as that under a H.C.:
\begin{equation}\label{5-24}
{\hat{\cal H}}({\bf x},t)\to{\hat{\cal H}}^\dag({\bf x},t)={\hat{\cal H}}({\bf x},t)
\end{equation}
The validity of both Eqs.(\ref{5-23}) and (\ref{5-24}) has been verified experimentally since the discovery of parity violation and the establishment (and development) of standard model in particle physics till today.

\section{VI. Summary and Discussion}
\label{sec:discussion}
\setcounter{equation}{0}
\renewcommand{\theequation}{6.\arabic{equation}}
\vskip 0.1cm

1. Since the CPT theorem was proved in the framework of QFT, the CPT invariance was often discussed at the level of QFT. By contrast, its discussion at the level of RQM was, to our knowledge, rarely seen. The reason might be as follows: After performing the CPT transformation on a particle's WF as Eq.(\ref{3-29}) at the level RQM, one encountered a WF with "negative energy" inevitably. And together with the "negative probability density", no concensus could be reached on its explanation within the physics community. In our opinion, the RQM cannot be considered complete until two sets of WFs and operators are introduced for particle versus antiparticle respectively and they are linked with the (newly defined) space-time inversion, Eq.(\ref{3-19}). Alternatively, the symmetry between particle and antiparticle can also be realized by the mass inversion ($m\to -m$) as shown in Eq.(\ref{3-25}) at the level of RQM (but not for QFT). The fact that the symmetry ${\cal P}{\cal T}={\cal C}$ exists in both RQM and QFT strongly hints that it is a basic postulate in physics rather than merely a consequence of the CPT theorem.

2. Evidently, the renamed CPT invariance, Eq.(\ref{5-1}), discovered by L\"{u}ders and Pauli is already there in the present framework of QFT. However, one thing is important here: We should assign the helicity ($h$) rather than the spin projection (in space) ($s$) to particle's state in the expansion of field operator, Eq.(\ref{5-10}), as required by the Eq.(\ref{5-1}) with Eq.(\ref{516}). The helicity of a particle is just opposite to that of its antiparticle after the operation Eq.(\ref{5-1}).

Therefore, the experimental tests for the CPT invariance should include not only the equal mass and lifetime of particle versus antiparticle, but also the following fact: A particle and its antiparticle with opposite helicities must coexist in nature with no exception. A prominent example is the neutrino --- A neutrino $\nu_L$ (antineutrino $\bar{\nu}_R$) is permanently left-handed (right-handed) polarized whereas the fact that no $\nu_R$ exists in nature must means no $\bar{\nu}_L$ as well (as verified by the GGS experiment \cite{24}).

3. Despite of Eqs.(\ref{3-30}) and (\ref{5-1}) being always valid, we wish to stress that all discussions about C, P, CP and T symmetries remain meaningful regardless of being conserved or not. In particular, T always means the "reversal of motion" at the level of either QM or QFT even though it is always an antiunitary operator.

4. In hindsight, after learning Feshbach-Villars dissociation  of KG equation Eqs.(\ref{54})-(\ref{55}), we realized that the postulate, \ie, the renamed CPT invariance Eqs.(\ref{3-30}) and (\ref{5-1}) could reflect some intrinsic property of the theory of SR established by Einstein in 1905.

Actually, there are two Lorentz invariants in the kinematics of SR:
\begin{equation}\label{6-1}
c^2(t_1-t_2)^2-({\bf x}_1-{\bf x}_2)^2=c^2(t'_1-t'_2)^2-({\bf x'}_1-{\bf x'}_2)^2=const
\end{equation}
\begin{equation}\label{6-2}
E^2-c^2{\bf p}^2=E'^2-c^2{\bf p'}^2=m^2c^4
\end{equation}
It seems quite clear that Eq.(\ref{6-1}) is invariant under the space-time inversion (${\bf x}\to-{\bf x},t\to-t$) and Eq.(\ref{6-2}) remains invariant under the mass inversion ($m\to-m$). We believe that these two discrete symmetries are deeply rooted at the SR's dynamics via its combination with QM and developing into RQM and QFT --- the particle and its antiparticle are treated on equal footing and linked by the symmetry ${\cal P}{\cal T}={\cal C}$ essentially.

Hence, the invariance of a theory (either in RQM or in classical physics) under the space-time inversion or the mass inversion in one coordinate frame could be used as a tool to find or test a new equation for it being relativistic or not\cite{11,12,13,35,38,39,40}.

\section*{Appendix A: Klein Paradox for Klein-Gordon Equation and Dirac Equation}

We will discuss the Klein paradox for both KG equation and Dirac equation based on the basic postulate, Eq.(\ref{3-30}), at the level of QM.

\subsection*{AI: Klein Paradox for KG Equation}

Consider that a KG particle moves along $z$ axis in one-dimensional space and hits a step potential
\begin{equation*}
V(z)=\left\{
       \begin{array}{ll}
         0, & \hbox{$z<0$;} \\
         V_0, & \hbox{$z>0$.}
       \end{array}
     \right.\eqno{(A.1)}
\end{equation*}
Its incident WF with momentum $p\,(>0)$ and energy $E\,(>0)$ reads
\begin{equation*}
\psi_i=a\exp[i(pz-Et)],\quad (z<0)\eqno{(A.2)}
\end{equation*}
If $E=\sqrt{p^2+m^2}<V_0$, we expect that the particle wave will be partly reflected at $z=0$ with WF $\psi_r$ and another transmitted wave $\psi_t$ emerged at $z>0$:
\begin{equation*}
\psi_r=b\exp[i(-pz-Et)],\quad (z<0)\eqno{(A.3)}
\end{equation*}
\begin{equation*}
\psi_t=b'\exp[i(p'z-Et)],\quad (z>0)\eqno{(A.4)}
\end{equation*}
with $p'^2=(E-V_0)^2-m^2$. See Fig.1(a).
\begin{figure}
  \hspace*{-3mm}\includegraphics[scale=0.98]{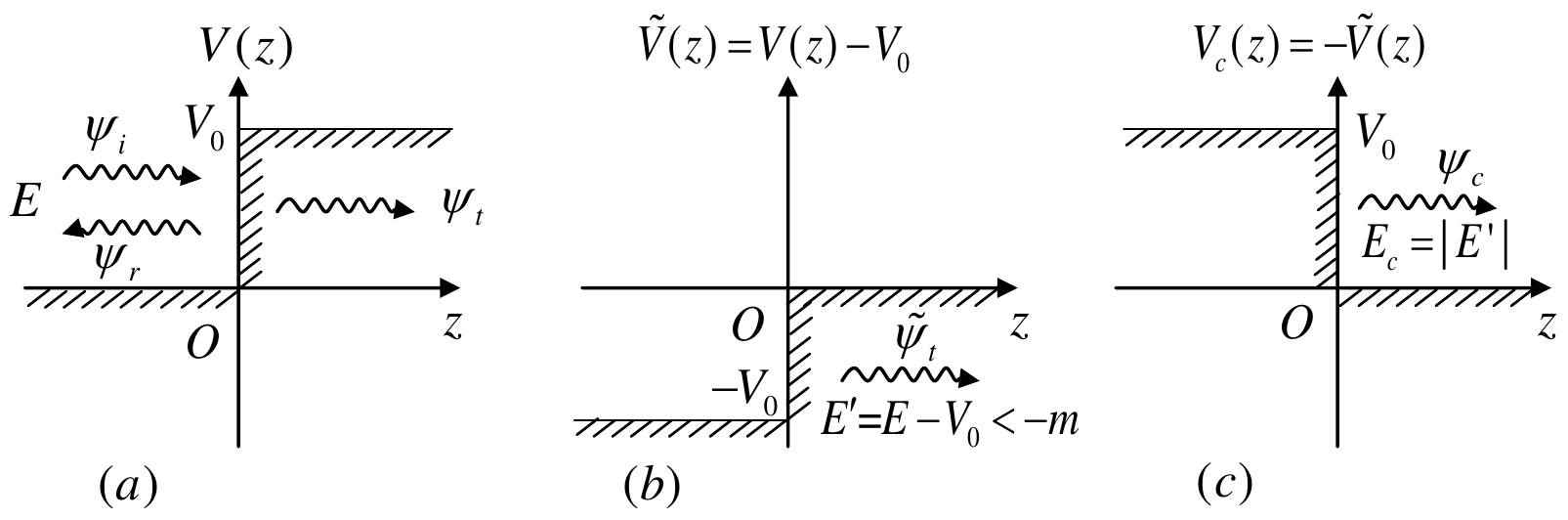}\\
  \caption{Klein paradox: (a) If $V_0>E+m$, there will be a wave $\psi_t$ at $z>0$.\\ (b) Just look at $z>0$ region, making a shift $V(z)\to\tilde{V}(z)=V(z)-V_0,\,E\to E'=E-V_0<-m$.\\ (c) An antiparticle (at $z>0$) appears with its energy $E_c=|E'|>m$ and the potential is $V_c(z)=-\tilde{V}(z)$}
\end{figure}

Two continuity conditions for WFs and their space derivatives at the boundary $z=0$ give two simple equations
\begin{equation*}
\left\{\begin{array}{l}
         a+b=b' \\
        (a-b)p=b'p'
       \end{array}\right.
\eqno{(A.5)}
\end{equation*}
The Klein paradox happens when $V_0>E+m$ because the momentum $p'=\pm\sqrt{(V_0-E)^2-m^2}$ is real again and the reflectivity $R$ of incident wave reads
\begin{equation*}
R=\left|\dfrac{b}{a}\right|^2=\left|\dfrac{p-p'}{p+p'}\right|^2,\;\left\{\begin{array}{l}
         R<1,\quad\text{if}\;p'>0 \\
        R>1,\quad\text{if}\;p'<0
       \end{array}\right.
\eqno{(A.6)}
\end{equation*}
(See Ref.\cite{6} or \S9.4 in Ref.\cite{11}, where discussions were not complete and need to be complemented and corrected here). Because the kinetic energy $E'$ at $z>0$ is negative: $E'=E-V_0<0$, what does it mean? Does the particle still remain as a particle?

As discussed in section IV, for a KG particle (or its antiparticle), two criterions must be held: its probability density $\rho$ (or $\rho_c$) must be positive and its probability current density ${\bf j}$ (or ${\bf j}_c$) must be in the same direction of its momentum ${\bf p}$ (or ${\bf p}_c$).

See Fig.1(b), after making a shift in the energy scale, \ie, basing on the new vacuum at $z>0$ region, we redefine a WF $\tilde{\psi}_t$ (which is actually the WF in the "interaction picture", $\tilde{\psi}_t=\psi_te^{iV_0t}\,(z>0)$)
\begin{equation*}
\psi_t\to\tilde{\psi}_t=b'\exp[i(p'z-E't)],\quad (z>0)\eqno{(A.7)}
\end{equation*}
($E'=E-V_0<0$). From now on we will replace KG WF $\tilde{\psi}_t$ by $\tilde{\phi}_t$ and $\tilde{\chi}_t$ according to Eq.(\ref{55}), if $\tilde{\psi}_t$ still describes a "particle", whose probability density $\rho_t$ should be evaluated by Eq.(\ref{56}) with $V\to\tilde{V}(z)=0\,(z>0)$ yielding:
\begin{equation*}
\rho_t=|\tilde{\phi}_t|^2-|\tilde{\chi}_t|^2=\dfrac{E'}{m}|b'|^2<0,\quad (z>0)\eqno{(A.8)}
\end{equation*}
And its probability current density $j_t$ should be given by Eq.(\ref{53}), yielding:
\begin{equation*}
j_t=\dfrac{p'}{m}|b'|^2,\quad (z>0)\eqno{(A.9)}
\end{equation*}
Eq.(A.8) is certainly not allowed. So to consider a "particle" with momentum $p'>0$ moving to the right makes no sense. Instead, we should consider $p'<0$ (which also makes no sense for a particle due to the boundary condition) and regard $\tilde{\psi}_t$ as an antiparticle's WF by rewriting it as:
\begin{equation*}
\tilde{\psi}_t=\psi_c=b'\exp[-i(p_cz-E_ct)],\quad (z>0)\eqno{(A.10)}
\end{equation*}
Now using Eq.(\ref{17}) we see that Eq.(A.10) does describe an antiparticle with momentum $p_c=-p'=|p'|=\sqrt{E_c^2-m^2}>0$ and energy $E_c=|E'|=V_0-E>0$. In the mean time, from the antiparticle's point of view (\ie, with $E_c>m$), the potential becomes $V_c(z)=-\tilde{V}(z)$ (comparing Eq.(\ref{55}) with Eq.(\ref{3-17})) as shown by Fig.1(c).

It is easy to see from Eqs.(\ref{3-16}),(\ref{3-18}) and (A.10) that\footnotemark[1]\footnotetext[1]{We had discarded the solution of $p'>0$ in Eqs.(A.8)-(A.9) as a particle. However, if we consider $p'=-p_c>0$ for an antiparticle, then similar to Eqs.(A.10)-(A.11), we would get $\rho_t^c>0$ but both $j_t^c$ and $p_c$ are negative, meaning that the antiparticle is coming from $z=\infty$, not in accordance with our boundary condition. So the case of $p'>0$ should be abandoned either as a particle or as an antiparticle.}
\begin{equation*}
\left\{\begin{array}{l}
       \rho_t^c=|\tilde{\chi}_t^c|^2-|\tilde{\phi}_t^c|^2=\dfrac{E_c}{m}|b'|^2>0,   \\[4mm]
       j_t^c=\dfrac{p_c}{m}|b'|^2
       \end{array}\right.
\quad (z>0)\eqno{(A.11)}
\end{equation*}
So the reflectivity, Eq.(A.6), should be fixed as:
\begin{equation*}
R_{KG}=\left|\dfrac{b}{a}\right|^2=\left|\dfrac{p+p_c}{p-p_c}\right|^2=\left(\dfrac{1+\gamma'}{1-\gamma'}\right)^2,\;
\gamma'=\dfrac{p_c}{p}>0 \eqno{(A.12)}
\end{equation*}
And the transmission coefficient can also be predicted as:
\begin{equation*}
T_{KG}=\dfrac{j_t^c}{j_i}=\dfrac{p_c}{p}\left|\dfrac{b'}{a}\right|^2=\dfrac{p_c}{p}\left|1+\dfrac{b}{a}\right|^2
=\dfrac{4pp_c}{(p-p_c)^2}=\dfrac{4\gamma'}{(1-\gamma')^2}\eqno{(A.13)}
\end{equation*}
\begin{equation*}
 R_{KG}-T_{KG}=1\eqno{(A.14)}
\end{equation*}
The variation of $T_{KG}$ seems very interesting:
\begin{equation*}
T_{KG}=\left\{\begin{array}{l}
0,\;\gamma'\to 0\quad(p_c\to 0,E_c\to m) \\[3mm]
\infty,\;\gamma'\to 1\quad(p_c=p,E_c=E=V_0/2)\\[3mm]
0,\;\gamma'\to \infty\quad(p_c\to \infty,E_c=V_0-E\to \infty)\\[3mm]
0,\;\gamma'\to \infty\quad(p\to 0,E\to m)
              \end{array}\right.
\eqno{(A.15)}
\end{equation*}
Above equations show us that the incident KG particle triggers a process of "pair creation" occurring at $z=0$, creating new particles moving to the left side (to join the reflected incident particle) so enhancing the reflectivity $R_{KG}>1$ and new antiparticles (with equal number of new particles) moving to the right.

To our understanding, this is not a stationary state problem for a single particle, but a nonstationary creation process of many particle-antiparticle system. It is amazing to see the Klein paradox in KG equation being capable of giving some prediction for such kind of process at the level of RQM. Further investigations are needed both theoretically and experimentally. \footnotemark[2]\footnotetext[2]{We find from the Google search that R. G. Winter in 1958 had examined the Klein paradox for KG equation at the QM level and reached basically the same result as ours. So he was the first author dealing with this problem. Regrettably, it seems that his paper had never been published on some journal.}

\subsection*{AII: Klein Paradox for Dirac Equation}

Beginning from Klein \cite{27}, many authors \eg Greiner \etal \cite{36,37}, have studied this topic. We will join them by using the similar approach like that for KG equation discussed above.

Based on similar picture shown in Fig.1, now we have three Dirac WFs under the condition $V_0>E+m$:
\begin{equation*}
\psi_i=a\begin{pmatrix}1\\0\\\frac{p}{E+m}\\0\end{pmatrix}
e^{i(pz-Et)},\psi_r=b\begin{pmatrix}1\\0\\\frac{-p}{E+m}\\0\end{pmatrix}
e^{i(-pz-Et)}\quad (z<0)\eqno{(A.16)}
\end{equation*}
\begin{equation*}
\psi_t=b'\begin{pmatrix}1\\0\\\frac{p'}{E-V_0+m}\\0\end{pmatrix}e^{i(p'z-Et)}
=b'\begin{pmatrix}1\\0\\\frac{-p'}{V_0-E-m}\\0\end{pmatrix}e^{i(p'z-Et)}
=\begin{pmatrix}\phi_t\\\chi_t\end{pmatrix}\quad (z>0)\eqno{(A.17)}
\end{equation*}
where $p'=\pm\sqrt{(V_0-E)^2-m^2}$. Unlike Eq.(A.8) for KG equation, the probability density for Dirac WF $\psi_t$ is positive definite (see Eq.(\ref{81}))
\begin{equation*}
\rho_t=\psi^\dag_t\psi_t=\phi^\dag_t\phi_t+\chi^\dag_t\chi_t\eqno{(A.18)}
\end{equation*}
Hence we will rely on two criterions: First, the probability current density and momentum must be in the same direction for either a particle or antiparticle. For $\psi_i$ and $\psi_r$, their probability current density are ($c=1$)
\begin{equation*}\begin{array}{l}
j_i=\psi^\dag_i\alpha_z\psi_i=\phi_i^\dag\sigma_z\chi_i+\chi^\dag_i\sigma_z\phi_i=\dfrac{2p}{E+m}|a|^2>0 \\
j_r=\psi^\dag_r\alpha_z\psi_r=\dfrac{-2p}{E+m}|b|^2<0
                 \end{array}\quad (z<0)
\eqno{(A.19)}
\end{equation*}
as expected. However, for $\psi_t$, we meet difficulty similar to that in Eq.(A.9)
\begin{equation*}
j_t=\psi^\dag_t\alpha_z\psi_t=\dfrac{-2p'}{V_0-E-m}|b'|^2 \quad (z>0)\eqno{(A.20)}
\end{equation*}
the direction of $j_t$ is always opposite to that of $p'$! The second criterion is: while $|\phi|>|\chi|$ for particle, we must have $|\chi_c|>|\phi_c|$ for antiparticle. Now in $\psi_i$ (or $\psi_r$), $|\phi_i|>|\chi_i|$ (or $|\phi_r|>|\chi_r|$), but the situation in $\psi_t$ is dramatically changed, the existence of $V_0$ renders $|\chi_t|>|\phi_t|$!

The above two criterions, together with the experience in KG equation, prompt us to choose $p'<0$ and regard $\psi_t$ as an antiparticle's WF. So we rewrite:
\begin{equation*}
\psi_t=\psi_t^ce^{-iV_0t}\eqno{(A.21a)}
\end{equation*}
\begin{equation*}
\psi_t^c=b'\begin{pmatrix}1\\0\\\frac{p_c}{E_c-m}\\0\end{pmatrix}
e^{-i(p_cz-E_ct)}=\begin{pmatrix}\phi_t^c\\\chi_t^c\end{pmatrix},
\tilde{\psi}^c_t=b'_c\begin{pmatrix}1\\0\\\frac{p_c}{E_c+m}\\0\end{pmatrix}
e^{-i(p_cz-E_ct)}=\begin{pmatrix}\chi_t^c\\\phi_t^c\end{pmatrix}
\quad (z<0)\eqno{(A.21b)}
\end{equation*}
where $\tilde{\psi}^c_t=(-\gamma^5)\psi_t^c$ (with new normalization constant $b'_c$ replacing
$b'$) describes an antiparticle with momentum $p_c=|p'|=-p'=\sqrt{E_c^2-m^2}>0$, energy $E_c=V_0-E>0$ and $|\chi_t^c|>|\phi_t^c|$. Using Eq.(\ref{82}) we find
\begin{equation*}
j_t^c=\dfrac{2p_c}{E_c+m}|b'_c|^2>0,\quad (z>0)\eqno{(A.22)}
\end{equation*}
as expected. Now it is easy to match Dirac WFs at the boundary $z=0$, ($\psi_i+\psi_r)|_{z=0}=\tilde{\psi}_t^c|_{z=0}$, yielding\footnotemark[1]\footnotetext[1]{Eq.(A.23) means that the large (small) component of spinor is connected with
large (small) component at both sides of $z=0$. However, if instead of $\tilde{\psi}^c_t$, the $\psi_t^c$ is used directly with its first (small) component being connected with the first (large) components of $\psi_i$ and $\psi_r$, it would lead to a different expression of Eq.(A.27): $\gamma\to\tilde{\gamma}=\sqrt{\frac{(E_c-m)(E-m)}{(E+m)(E_c+m)}}$, which is just the $1/\gamma$ ($\gamma$ and $1/\gamma$ make no difference in the result of, say, Eqs.(A.24) and (A.25)) defined by Eq.(8) on page 266 of Ref.\cite{36} (see Eq.(A31) below) or that by Eq.(5.36) in Ref.\cite{37}}
\begin{equation*}\left\{\begin{array}{l}
a+b=b'_c \\
\dfrac{(a-b)p}{E+m}=\dfrac{b'_cp_c}{E_c+m}
                 \end{array}\right.\to\left\{\begin{array}{l}
\dfrac{b}{a}=\dfrac{\xi-\eta}{\xi+\eta} \\
\dfrac{b'_c}{a}=1+\dfrac{b}{a}=\dfrac{2\xi}{\xi+\eta}
                 \end{array}\right.
\eqno{(A.23)}
\end{equation*}
where $\xi=p(E_c+m)>0,\eta=p_c(E+m)>0$. The reflectivity $R_D$ and transmission coefficient $T_D$ follow from Eq.(A.19) and (A.22) as:
\begin{equation*}
R_D=\dfrac{|j_r|}{j_i}=\left|\dfrac{b}{a}\right|^2=\left(\dfrac{1-\gamma}{1+\gamma}\right)^2\eqno{(A.24)}
\end{equation*}
\begin{equation*}
T_D=\dfrac{j_t^c}{j_i}=\left|\dfrac{b'_c}{a}\right|^2\dfrac{p_c(E+m)}{p(E_c+m)}=\dfrac{4\gamma}{(1+\gamma)^2}\eqno{(A.25)}
\end{equation*}
\begin{equation*}
 R_D+T_D=1\eqno{(A.26)}
\end{equation*}
where
\begin{equation*}
\gamma=\dfrac{\eta}{\xi}=\sqrt{\dfrac{(E_c-m)(E+m)}{(E-m)(E_c+m)}}\geq0\;(E_c=V_0-E\geq m)\eqno{(A.27)}
\end{equation*}
and
\begin{equation*}
T_D=\left\{\begin{array}{l}
0,\;\gamma\to 0\quad(p_c\to 0,E_c\to m) \\
1,\;\gamma= 1\quad(p_c=p,E_c=E=V_0/2)\;(\text{resonant\ transmission}) \\
\frac{2p}{E+p},\;\gamma\to \sqrt{\frac{E+m}{E-m}}\quad(E_c=V_0-E\to \infty)\\
0,\;\gamma\to \infty\quad(p\to 0,E\to m)
              \end{array}\right.
\eqno{(A.28)}
\end{equation*}
The variation of $T_D$ bears some resemblance to Eq.(A.15) for KG equation but shows striking difference due to sharp contrast between Eqs.(A.24)-(A.28) and Eqs.(A.12)-(A.15).

To our understanding, in the above Klein paradox for Dirac equation, there is no "pair creation" process occurring at the boundary $z=0$. The paradox just amounts to a steady transmission of particle's wave $\psi_i$ into a high potential barrier $V_0>E+m$ at $z>0$ region where $\psi_t$ shows up as an antiparticle's WF propagating to the right. In some sense, the existence of a potential barrier $V_0$ plays a "magic" role of transforming the
particle into its antiparticle. Because the probability densities of both particle and antiparticle are positive definite, the total probability can be normalzed over the entire space like that for one particle case:
\begin{equation*}
\int_{-\infty}^\infty[\rho(z)\Theta(-z)+\rho_c(z)\Theta(z)]dz=1\eqno{(A.29)}
\end{equation*}
($\Theta(z)$ is the Heaviside function) and the probability current density remains continuous at the boundary $z=0$. In other words, the continuity equation holds in the whole space just like what happens in a one-particle stationary state.

It is interesting to compare our result with that in Refs.\cite{36} and \cite{37}. In Ref.\cite{36}, Eqs.(13.24)-(13.28) are essentially the same as ours. But the argument there for choosing $\bar{p}<0$ in Eq.(13.23) is based on the criterion of the group velocity $v_{gr}$ being positive (for the transmitted wave packet moving toward $z=\infty$).  And the $v_{gr}$ is stemming from Eq.(13.16) which is essentially the probability current density in our Eq.(A.19) or (A.20).

However, the author in Ref.\cite{36} also considered the other choice $\bar{p}>0$ in the example (p.265-267 in \cite{36}) based on the hole theory, ending up with the prediction as:
\begin{equation*}
R=\left(\dfrac{1+\gamma}{1-\gamma}\right)^2,\;T=\dfrac{4\gamma}{(1-\gamma)^2},\;R-T=1\eqno{(A.30)}
\end{equation*}
where
\begin{equation*}
\gamma=\dfrac{p_2}{p_1}\dfrac{E+m}{V_0-E-m}=\sqrt{\dfrac{(V_0-E+m)(E+m)}{(V_0-E-m)(E-m)}}\eqno{(A.31)}
\end{equation*}
The argument for the validity of his Eqs.(A.30)-(A.31) is based on the hole theory (see also section 5.2 in Ref.\cite{37}), saying that once $V_0>E+m$, there would be an overlap between the occupied negative continuum for $z>0$ and the empty positive continuum for $z<0$, providing a mechanism for electron-positron pair creation if the "hole" at $z>0$ can be identified with a positron. We doubt the "hole" theory seriously because there are only two electrons (with opposite spin orientations) staying at each energy level in the negative continuum. So it seems that there is no abundant source for electrons and "holes" to account for the huge value of $T>1$ in Eq.(A.30).

Fortunately, we learn from section 10.7 in Ref.\cite{37} that if the Klein paradox in Dirac equation is treated at the level of QFT, their result turns out to be the same form as our Eqs.(A.24)-(A.28), rather than Eqs.(A.30)-(A.31).

\section*{Acknowledgements}

\vskip 0.1cm

We thank E. Bodegom, T. Chang, Y. X. Chen, T. P. Cheng, X. X. Dai, G. Tananbaum, V. Dvoeglaznov, Y. Q. Gu, F. Han, J. Jiao, A. Kellerbauer, A. Khalil, R. Konenkamp, D. X. Kong, J. S. Leung, P. T. Leung, Q. G. Lin, S. Y. Lou, D. Lu, Z. Q. Ma, D. Mitchell, E. J. Sanchez, Z. Y. Shen, Z. Q. Shi, P. Smejtek, X. T. Song, R. K. Su, Y. S. Wang, Z. M. Xu, X. Xue, J. Yan, F. J. Yang, J. F. Yang, R. H. Yu, Y. D. Zhang and W. M. Zhou for encouragement, collaborations and helpful discussions.




\begin{thebibliography}{99}

\bibitem[1]{1}
A. Einstein, B. Podolsky and N. Rosen, \pr{47}{1935}{777}.

\bibitem[2]{2}
A. Apostolakis \etal (CPLEAR Collaboration),, \plb{422}{1998}{339-348}.

\bibitem[3]{3}
D. Bohm, {\it Quantum Theory}, Prentice Hall, 1956.

\bibitem[4]{4}
J. S. Bell, {\it Physics}, Long Island City, NewYork, {\bf 1}, 195-200 (1964).

\bibitem[5]{5}
H. Guan, {\it Basic Concepts in Quantum Mechanics}, (High Education Press, Beijing, 1990), Chapter 7 (in Chinese).

\bibitem[6]{6}
G. J. Ni, H. Guan, W. M. Zhou and J. Yan, Chinese Phys. Lett. {\bf 17}, 393-395 (2000), {\tt quant-ph}/0001016.

\bibitem[7]{7}
O. Nachtman, {\it Elementary Particle Physics, Concepts and Phenomena}, Springer-Verlag, 1990.

\bibitem[8]{8}
W. Greiner and B. M\"{u}ller, {\it Gauge Theory of Weak Interactions}, Springer-Verlag, 1993, Ch.8.

\bibitem[9]{9}
E. J. Konopinski and H. M. Mahmaud, \pr{92}{1953}{1045}.

\bibitem[10]{10}
G. J. Ni and S. Q. Chen, Journal of Fudan University (Natural Science)
{\bf 35}, 325 (1996); English version is in "Photon and Poincar\'{e} Group",
edit. V. Dvoeglazov (NOVA Science Publisher, Inc. 1999), Chapter III, 145-169;
{\tt hep-th}/9508069.

\bibitem[11]{11}
G. J. Ni and S. Q. Chen, {\it Advanced Quantum Mechanics}, 2nd
Edition (Fudan University Press, 2003); English Edition was
published by Rinton Press, 2002.

\bibitem[12]{12}
G. J. Ni, in {\it Relativity, Gravitation, Cosmology}, Edit by V. V. Dvoeglazov and A. A. Espinoza Garrido,
NOVA Science Publisher, 2004, 123-136, \physics{0308038}.

\bibitem[13]{13}
G. J. Ni, J. J. Xu and S. Y. Lou, Chinese Physics {\bf B 20}, 020302 (2011); {\tt quant-ph}/0511197.

\bibitem[14]{14}
M. E. Peskin and D. V. Schroeder, {\it An Introdution to Quantum Field Theory}, Addison-Wesley Publishing Company, 1995.

\bibitem[15]{15}
J. J. Sakurai, {\it Advanced Quantum Mechanics}, Addison-Wesley Publishing Company, 1978.

\bibitem[16]{16}
J. J. Sakurai, {\it Modern Quantum Mechanics}, NewYork, John Wiley \& Sons, Inc. 1994.

\bibitem[17]{17}
J. D. Bjorken and S. D. Drell, {\it Relativistic Quantum Mechanics}, McGraw-Hill. 1964, {\it Relativistic Quantum Field}, McGraw-Hill. 1967.

\bibitem[18]{18}
T. D. Lee and C. N. Yang, \pr{104}{1956}{254}, \pr{105}{1957}1671.

\bibitem[19]{19}
C. S. Wu, E. Ambler, R. W. Hayward, D. D. Hoppes and R. P. Hudson, \pr{105}{1957}{1413}.

\bibitem[20]{20}
K. Nakamura \etal (Particle Data Group), Journal of Physics, G{\bf 37}, 075021(2010) or the Particle Physics Booklet extracted from it.

\bibitem[21]{21}
T. D. Lee and C. S. Wu, Ann. Rev. Nucl. Sci., {\bf 15}, 381-476, (1965).

\bibitem[22]{22}
G. J. Ni, Journal of Fudan University (Natural Science), 1974, No.3-4, 125 (In fact, this paper was in collaboration with Suqing Chen, but her name was erased according to editor's advice for promoting the publication at that time).

\bibitem[23]{23}
L. B. Okun, Phys. Today, {\bf 42}, 31-36 (June 1989); Discussions in
{\bf 42}, pp.13, 15, 115, 117 (May 1990)

\bibitem[24]{24}
M. Goldhaber, L. Grodgins and A. W. Sunyar, \pr{109}{1958}{1015}.

\bibitem[25]{25}
H. Feshbach and F. Villars, \rmp{30}{1958}{24}.

\bibitem[26]{26}
W. Pauli and V. Weisskopf, Phys. Acta. {\bf 7}, 709 (1934).

\bibitem[27]{27}
O. Klein, Z. Physik, {\bf 53}, 157 (1929).

\bibitem[28]{28}
G. J. Ni, Progress in Physics (Nanjing, China) {\bf 23}, 484-503 (2003), (In English).

\bibitem[29]{29}
E. C. G. St\"{u}eckelberg, Helv. Phys. Acta. {\bf 14}, 32L, 588 (1941).

\bibitem[30]{30}
R. P. Feynman, \pr{76}{1949}{749,769}.

\bibitem[31]{31}
G. L\"{u}ders, Dan. Mat. Fys. Medd. {\bf 28}, No.5 (1954).

\bibitem[32]{32}
G. L\"{u}ders, Ann. Phys. (N. Y.) {\bf 2}, 1-15 (1957).

\bibitem[33]{33}
W. Pauli, in {\it Niels Bohr and the Development of Physics}, ed. by W. Pauli, L. Rosenfeld and V. Weisskopf (McGraw-Hill, 1955), 30-51.

\bibitem[34]{34}
R. F. Streater and A. S. Weightman, {\it PCT, Spin and Statitics and All That}, W. A. Benjamin, 1964.

\bibitem[35]{35}
G. J. Ni, in {\it Relativity, Gravitation, Cosmology: New development}, Editor: V. Dvoeglazov, (NOVA Science Publisher, 2010), pp237-265.

\bibitem[36]{36}
W. Greiner, {\it Relativistic Quantum Mechanics}, Springer-Verlag, 1990, 261-267.

\bibitem[37]{37}
W. Greiner, B. M\"{u}ller and J. Rafelski, {\it Quantum Electrodynamics of Strong Fields}, Springer-Verlag, 1985.

\bibitem[38]{38}
T. Chang and G. J. Ni, FIZIKA (Zagreb), {\bf 11} (1), 49 (2002), {\tt hep-ph}/0009291.

\bibitem[39]{39}
G. J. Ni and T. Chang, Journal of Shaanxi Normal Univ. (Nat. Sci. Ed.), {\bf 30} (3), 32 (2002), {\tt hep-ph}/0103051.

\bibitem[40]{40}
G. J. Ni, {\it ibid.}, {\bf 29} (1), 1 (2001), {\tt hep-th}/0201077; {\bf 30} (4), 1 (2002), {\tt hep-th}/0203060.


\end{thebibliography}
\end{document}